\documentclass[traditabstract]{aa} 		

\usepackage{graphicx}
\usepackage{txfonts}
\usepackage{natbib}
\usepackage{enumitem}
\usepackage{lscape}

\begin{document}

\title{Panchromatic spectral energy distributions of \textit{Herschel} sources
\thanks{\textit{Herschel} is an ESA space observatory with science instruments provided by 
European-led Principal Investigator consortia and with important participation from NASA.}}

\author{S. Berta\inst{1}
        \and
	D. Lutz\inst{1}
	\and
	P. Santini\inst{2}
	\and
	S. Wuyts\inst{1}
	\and
	D. Rosario\inst{1}
	\and
	D. Brisbin\inst{3}
	\and
	A. Cooray\inst{4}
	\and
	A. Franceschini\inst{5}
	\and
	C. Gruppioni\inst{6}
	\and
	E. Hatziminaoglou\inst{7}
	\and
	H.~S. Hwang\inst{8}
	\and
	E. Le Floc'h\inst{9}
	\and
	B. Magnelli\inst{1}
	\and
	R. Nordon\inst{10}
	\and
	S. Oliver\inst{11}
	\and
	M.~J. Page\inst{12}
	\and
	P. Popesso\inst{1}
	\and
	L. Pozzetti\inst{6}
	\and
	F. Pozzi\inst{13}
	\and
	L. Riguccini\inst{14}
	\and
	G. Rodighiero\inst{5}
	\and
	I. Roseboom\inst{15}
	\and
	Douglas Scott\inst{16}
	\and
	M. Symeonidis\inst{12}
	\and
	I. Valtchanov\inst{17}
	\and
	M. Viero\inst{18} 
	\and 
	L. Wang\inst{11}
	}

\offprints{Stefano Berta, \email{berta@mpe.mpg.de}}

\institute{Max-Planck-Institut f\"{u}r Extraterrestrische Physik (MPE),
Postfach 1312, 85741 Garching, Germany.
\and
INAF - Osservatorio Astronomico di Roma, via di Frascati 33, 00040 Monte Porzio Catone, Italy.
\and
Center for Radiophysics and Space Research, Cornell University, Ithaca, NY 14853, USA.
\and
Dept. of Physics \& Astronomy, University of California, Irvine, CA 92697, USA.
\and
Dipartimento di Astronomia, Universit{\`a} di Padova, Vicolo dell'Osservatorio 3, 
35122 Padova, Italy.
\and
INAF-Osservatorio Astronomico di Bologna, via Ranzani 1, I-40127 Bologna, Italy.
\and
ESO, Karl-Schwarzschild-Str. 2, 85748 Garching bei M\"{u}nchen, Germany.
\and
Smithsonian Astrophysical Observatory, 60 Garden Street, Cambridge, MA 02138, USA.
\and
Laboratoire AIM, CEA/DSM-CNRS-Universit{\'e} Paris Diderot, IRFU/Service d'Astrophysique, 
B\^at.709, CEA-Saclay, 91191 Gif-sur-Yvette Cedex, France.
\and
School of Physics and Astronomy, The Raymond and Beverly Sackler
Faculty of Exact Sciences, Tel-Aviv University, Tel-Aviv 69978, Israel.
\and
Astronomy Centre, Department of Physics and Astronomy, University of Sussex, Brighton BN1 9QH.
\and
Mullard Space Science Laboratory, University College London, Holmbury St. Mary, Dorking, Surrey RH5 6NT.
\and
Dipartimento di Astronomia, Universit{\`a} di Bologna, Via Ranzani 1,
40127 Bologna, Italy.
\and
Astrophysics Branch, NASA -- Ames Research Center, MS 245-6, Moffett Field, CA 94035.
\and
Institute for Astronomy, University of Edinburgh, Royal Observatory, Blackford Hill, Edinburgh EH9 3HJ.
\and
Department of Physics \& Astronomy, University of British Columbia, 6224 Agricultural Road, Vancouver, BC V6T 1Z1, Canada.
\and 
Herschel Science Centre, ESAC, ESA, PO Box 78, Villanueva de la Can\~ada, 28691 Madrid, Spain.
\and
California Institute of Technology, 1200 E. California Blvd., Pasadena, CA 91125.
}

\date{Received 06/12/2012; accepted 16/01/2013}

\abstract{
Combining far-infrared \textit{Herschel} photometry from the PACS Evolutionary Probe (PEP) and 
\textit{Herschel} Multi-tiered Extragalactic Survey (HerMES) guaranteed time programs with 
ancillary datasets in the GOODS-N, GOODS-S, and COSMOS fields, it is possible to sample 
the 8--500\,$\mu$m spectral energy distributions (SEDs) of galaxies 
with at least 7--10 bands. Extending to the UV, optical, and near-infrared, the number of bands 
increases up to 43. We reproduce the distribution of galaxies in a carefully selected restframe 
ten colors space, 
based on this rich data-set, using a superposition of multi-variate Gaussian modes. 
We use this model to classify galaxies and build median SEDs 
of each class, which are then fitted with a modified version of the {\sc magphys} code that combines stellar light, 
emission from dust heated by stars and a possible warm dust contribution heated by an active galactic nucleus (AGN). 
The color distribution of galaxies in each of the considered fields can be well described
with the combination of 6--9 classes, spanning a large range of far- to near-infrared luminosity ratios, 
as well as different strength of the AGN contribution to bolometric luminosities.
The defined Gaussian grouping is used to identify rare or odd sources. The zoology of 
outliers includes \textit{Herschel}-detected ellipticals, very blue $z\sim1$ Ly-break galaxies, quiescent spirals, and 
torus-dominated AGN with star formation. 
Out of these groups and outliers, a new template library is assembled, consisting of 32 SEDs describing the 
intrinsic scatter in the restframe UV-to-submm colors of infrared galaxies.
This library is tested against $L(\textnormal{IR})$ estimates with and without \textit{Herschel} data included, and compared to 
eight other popular methods often adopted in the literature. When implementing \textit{Herschel} 
photometry, these approaches produce $L(\textnormal{IR})$ values consistent with each other within 
a median absolute deviation of 10--20\%, the
scatter being dominated more by fine tuning of the codes, rather than by the choice of SED templates.
Finally, the library is used to classify 24 $\mu$m detected sources in PEP GOODS fields on the basis 
of AGN content, $L(60)/L(100)$ color and $L(160)/L(1.6)$ luminosity ratio. 
AGN appear to be distributed in the stellar mass ($M_\ast$) vs. star formation rate (SFR) space along with all 
other galaxies, regardless of the amount of infrared luminosity they are powering,
with the tendency to lie on the high SFR side of the 
``main sequence''. The incidence of warmer star-forming 
sources grows for objects with higher specific star formation rates (sSFR), and they tend to populate the ``off-sequence'' 
region of the $M_\ast-\textnormal{SFR}-z$ space.}


\maketitle


\section{Introduction}\label{sect:intro}

The advent of wide-field surveys on the one hand, and deep pencil-beam 
observation of selected blank fields on the other, have produced, 
in recent years, extensive 
multi-wavelength photometric information for very large numbers of galaxies. 
Optical and near-infrared (NIR) surveys such as the Sloan Digital Sky Survey \citep[SDSS,][]{york2000}, the
Two Degree Field survey \citep{colless1999}, the Two Micron All Sky Survey
\citep[2MASS,][]{kleinmann1994}, the Cosmic Evolution Survey \citep[COSMOS,][]{scoville2007}, the Great Observatories 
Origins Deep Survey (GOODS, Dickinson et al. \citeyear{dickinson2001}),
the All-wavelength Extended Groth Strip International Survey \citep[Aegis,][]{davis2007} survey, the 
\textit{Spitzer} Wide-Area Infrared Extragalactic (SWIRE) Legacy
survey \citep{lonsdale2003,lonsdale2004}, among many others, 
include now millions of galaxies spanning across all epochs, to $z>4$ 
in the deepest cases, and covering the whole electromagnetic spectrum from 
X-rays to radio wavelengths.

At long wavelengths, extragalactic surveys with the 
Infrared Astronomical Satellite \citep[IRAS,][]{neugebauer1984}, 
Infrared Space Observatory \citep[ISO, e.g. ELAIS,][]{rowanrobinson2004}, 
\textit{Spitzer} \citep[e.g.][Dickinson et al. 2001]{frayer2009,lefloch2009,magnelli2009,lonsdale2003,lonsdale2004}
and AKARI \citep[e.g.][]{matsuura2009,matsuhara2006}
revealed the mid- and far-infrared (MIR, FIR) counterparts of local galaxies first, and 
of several thousands of distant sources up to $z\sim3$ in the latest incarnation 
of 80-cm class IR space telescopes. 

While detailed spectroscopic studies are still very demanding in terms 
of observing time, and include a relatively limited number of sources, 
especially at $z>1.0-1.5$ and at the faint end of the luminosity function, 
a variety of techniques has been developed to extract the wealth of information 
stored in photometric data. 
Fitting the broad band spectral energy distribution (SED) of each source by 
adopting a library of galaxy templates can provide a classification, together with 
an estimate of luminosity (in the desired bands), stellar mass 
and other possible parameters. However, success depends
on the predefined templates being representative of the galaxies
under consideration.

Basic approaches to SED fitting follow different paths, depending 
on goals and wavelength coverage. Evolutionary and mixed
stellar population synthesis, including global or age-selective 
extinction corrections, has proven to be very successful 
in the optical and near-infrared 
\citep[e.g.][]{bruzual2003,poggianti2001,bressan1994,renzini1986}
to study galaxy properties such as stellar mass, star formation rate and history, 
and has been sometimes extended to longer wavelengths by means of energy balance
arguments \citep[e.g.][]{dacunha2008,berta2004}. Modeling 
from the ultraviolet (UV) to FIR, making use of radiative transfer and ray
tracing techniques \citep[e.g.][]{siebenmorgen2007,piovan2006,dopita2005,efstathiou2000,silva1998},
provides extensive information on the physical properties of 
star-forming regions, molecular clouds, and the diffuse dust component, as 
well as of stellar populations. Finally semi-empirical libraries 
(e.g. Polletta et al. \citeyear{polletta2007}; Kirkpatrick et al. 
\citeyear{kirkpatrick2012}; and this work) offer flexible 
templates to be compared in a fast manner to large observed catalogs 
for quick classification and derivation of basic properties.

The semi-empirical library built by \citet{polletta2007} spans from the UV to submm, 
and covers a wide range of spectral types, from elliptical galaxies to 
spirals and irregulars, and includes templates of 
prototype starbursts, ultra-luminous infrared galaxies (ULIRGs), 
and objects with different amounts of an active galactic nucleus (AGN) power (both type-1 and type-2).

Alternatives in the infrared domain focus on 
star-forming objects and are based on locally determined properties such as 
the local luminosity-temperature relation, implying that a single SED shape is associated with each value of $L(\textnormal{IR})$   
 \citep[e.g.][the latter two already pointing out weaknesses of this approach]{chary2001,dale2002,elbaz2010,nordon2010}.
Recent works indicate that the SED shape of infrared galaxies is
more naturally explained in relation to their distance from the 
so-called ``main sequence'' of star formation (Elbaz et al. \citeyear{elbaz2011},
Nordon et al. \citeyear{nordon2012}; see also Magnelli et al., in prep.). 
Defined in the $M_\ast-\textnormal{SFR}$ plane, this main sequence (MS)
represents the ``secular'' and dominant mode
of baryon transformation into stars 
\citep[e.g.][]{elbaz2011,rodighiero2011,wuyts2011b,daddi2009,daddi2007a,elbaz2007},
likely supported by accretion of gas from the intergalactic medium 
\citep[e.g.][and references therein]{dave2010}. In this picture, galaxies lying above 
the MS would be undergoing powerful, starbursting events,
possibly triggered by major mergers.  

The above-mentioned SED libraries, as well as others not listed, 
have not only been employed to derive properties of 
individual sources, but have also been widely used in describing 
the statistical properties of galaxy populations, for example 
while interpreting and modeling observed number counts
\citep[e.g.][]{berta2011,berta2010,bethermin2011b,franceschini2010}
or assessing luminosity functions at different wavelengths \citep[e.g.][]{gruppioni2013,gruppioni2010,rodighiero2010b}.

Most of the templates adopted in the literature 
have actually been built using local galaxies 
\citep[e.g.][]{rieke2009,polletta2007,dale2002,chary2001}, and 
although they have proven to reproduce the SEDs of
high-$z$ galaxies reasonably well, their applicability can be subject to
criticism. \citet{gruppioni2010} have shown, for example, 
that the observed SEDs of \textit{Herschel} detected galaxies 
are in general well reproduced by the \citet{polletta2007} 
library, but very bright FIR emitters, or objects 
intermediate between spirals and starburst, are actually poorly 
represented. 

After three years of operation, the \textit{Herschel} Space Observatory \citep{pilbratt2010} 
has proven to be the ultimate machine to build 
detailed FIR SEDs of galaxies up to $z>3$, without being limited to 
very luminous sources. 
The sensitivity reached by 
the Photodetector Array Camera and Spectrometer \citep[PACS,][]{poglitsch2010} and 
the Spectral and Photometric Imaging REceiver \citep[SPIRE,][]{griffin2010}
instruments aboard \textit{Herschel} has guaranteed the detection not only of 
the most extreme star-forming galaxies, with 
star formation time scales on the order of $\sim0.1$ Gyr or less \citep[e.g.][]{magnelli2010,magnelli2012,rodighiero2011},
but also objects lying on the MS of star formation up to $z\sim2$.
In addition, the variety of \textit{Herschel} detections includes objects with MIR power-law spectra, dominated by an 
AGN torus emission, but with optical stellar SEDs \citep[e.g.][]{hatziminaoglou2010}, as well as type-1 AGN.

This paper is based on \textit{Herschel} FIR observations carried out 
in the framework of the PACS Evolutionary Probe \citep[PEP,][]{lutz2011} and the
\textit{Herschel} Multi-tiered Extragalactic Survey \citep[HerMES,][]{oliver2012} in three of the 
most popular extragalactic ``blank fields'': GOODS-N, GOODS-S, and COSMOS (Dickinson et al. \citeyear{dickinson2001}; 
Scoville et al. \citeyear{scoville2007}).
Combining \textit{Herschel} and \textit{Spitzer} data, it is now possible to sample 
the 8--500 $\mu$m SEDs of galaxies 
with 7 to 10 bands, thus accurately probing the shape 
of the dust emission peak, as well as MIR light dominated by warmer dust.
Extending to the UV, optical, and NIR, the number of bands 
increases up to 43 in the most favorable case (COSMOS, including intermediate bands).

We take advantage of this rich multi-wavelength data-sets to build a new set of 
semi-empirical templates, based on the observed UV to FIR photometry 
of galaxies detected by \textit{Herschel} (but not only), without any restriction in 
redshift.

The objects are going to be grouped on the basis of their UV-FIR restframe 
colors: following the inspiring work of \citet{davoodi2006}, we model the multi-color distribution 
of galaxies with a mixture of multivariate Gaussians \citep{connolly2000,nichol2001b}.
This technique is independent of the predetermined galaxy template libraries,
and provides the identification, classification, and characterization of
both existing and new object types, a compact
description of the data, and a recipe for the identification of outliers.
Each source is then classified as belonging to one of these Gaussian modes, 
and a median SED is derived for all identified groups. 

These SEDs are going to be fitted using a custom, modified version of {\sc magphys} 
\citep{dacunha2008}, combining 
stellar light, emission from dust heated by stellar populations and 
a possible AGN contribution.
Our main product is a robust library of templates
which can cope with the very large scatter in observed colors of infrared galaxies 
over the full UV, optical, NIR, MIR and FIR spectral range, thus 
discriminating between different classes of galaxies, including AGN-harboring objects, 
and other obscured sources.

This paper is structured as follows. Section \ref{sect:data} describes the available 
data-sets and the simple procedure adopted to derive restframe colors; Sect. \ref{sect:FastEM}
deals with the classification scheme, based on the parametric description of 
the multi-variate distribution in the chosen $N$-dimensional color space. 
Interpolation of median SEDs by means of a multi-component fit is presented in 
Sect. \ref{sect:sed_fit}. In Sect. \ref{sect:outliers} we use the results of 
the previous decomposition of the restframe multi-color distribution to identify 
outliers and very rare sources.
Finally, Sect. \ref{sect:discussion} applies the new library to the derivation of 
infrared luminosities, comparing results to eight other popular methods often adopted in 
the literature. The newly derived UV-to-submm templates are used to classify 
MIPS 24~$\mu$m detected objects in the GOODS fields.
Throughout our analysis, we adopt a standard $\Lambda$CDM cosmology 
with $H_0=71$ $[$km s$^{-1}$ Mpc$^{-1}]$, $\Omega_\textnormal{m}=0.27$, and $\Omega_\Lambda=0.73$.


\section{Building color catalogs}\label{sect:data}

The PACS Evolutionary Probe survey
(PEP\footnote{http://www.mpe.mpg.de/ir/Research/PEP/}, Lutz et al. 
\citeyear{lutz2011}) observed 
the most popular and widely studied extragalactic blank fields 
(The Lockman Hole, EGS, ECDFS, COSMOS, GOODS-N, and GOODS-S) with the PACS \citep{poglitsch2010} 
instrument aboard \textit{Herschel} \citep{pilbratt2010} at 70, 100 and 160 $\mu$m.
In parallel, the HerMES survey \citep{oliver2012} covered the same (and additional) 
fields at longer wavelengths (250, 350, 500 $\mu$m) with the SPIRE instrument 
\citep{griffin2010}. We first describe the multi-wavelength data-sets available 
in the adopted fields, and then enter into the details of source selection, 
which is mostly driven by the need to derive restframe colors of galaxies.

\subsection{Available data}

Here we exploit PEP and HerMES data in the Great Observatories Origins Deep Survey 
(GOODS, Dickinson et al. \citeyear{dickinson2001}) northern and southern fields, as well as in COSMOS \citep{scoville2007}.
\textit{Herschel} data are combined with ancillary catalogs covering the electromagnetic spectrum 
from UV to FIR wavelengths. Table \ref{tab:depths} summarizes 
\textit{Herschel} depths reached in the three fields considered.

We defer to dedicated publications for details on reduction, map making and 
simulations for PACS \citep{lutz2011,popesso2012,berta2010} and 
SPIRE \citep{levenson2010,viero2012} datasets.

\textit{Herschel} photometry was performed through point spread function (PSF) fitting, adopting 
\textit{Spitzer} MIPS 24 $\mu$m detected sources as positional priors. 
This approach maximizes depth, optimizes deblending, and facilitates 
the match of the FIR photometry to ancillary data, because 
24 $\mu$m catalogs are linked to \textit{Spitzer}/IRAC (3.6 to 8.0 $\mu$m) 
and then optical bands, either through 
prior extraction (e.g. see Magnelli et al. \citeyear{magnelli2009} for GOODS 
fields) or PSF-matching (e.g. see Berta et al. \citeyear{berta2011}).
PACS priors source extraction followed the method described by \citet{magnelli2009}, 
while SPIRE fluxes were obtained with the \citet{roseboom2010} recipe. 
Images at all \textit{Herschel} wavelengths have undergone extractions using the same MIPS prior catalogs, i.e. 
those presented by \citet{magnelli2009,magnelli2011} in the GOODS fields and 
by \citet{lefloch2009} in COSMOS.

In GOODS-S we make use of the MUSIC \citep{grazian2006} photometric catalog, 
with the addition of GALEX and \textit{Spitzer} IRS 16 $\mu$m \citep{teplitz2011} data,
matched to the 24--500 $\mu$m catalog via a simple closest neighbor algorithm. 
Spectroscopic redshifts were collected for more than 3000 sources
\citep[see][and references therein]{berta2011}.

A PSF-matched catalog\footnote{The multi-wavelength GOODS-N catalog is available on the 
PEP web page at the URL http://www.mpe.mpg.de/ir/Research/PEP/public\_data\_releases.php.} was created in GOODS-N, including
photometry from GALEX UV to \textit{Spitzer}, IRAC and MIPS 24 $\mu$m 
\citep[see][for further details]{berta2011,berta2010}. This catalog includes
the collection of spectroscopic redshifts by \citet{barger2008} and 
new photometric redshifts derived with the {\sc eazy} \citep{brammer2008} code.
To these, we add the \textit{Spitzer} IRS 16 $\mu$m photometry by \citet{teplitz2011}.

Finally, in COSMOS we adopt the public\footnote{http://irsa.ipac.caltech.edu/data/COSMOS/}
UV-optical catalog, including $U$ to $K$ broad and intermediate bands \citep{capak2007,ilbert2009}, 
combined with public spectroscopic \citep{lilly2009,trump2009} redshifts. New photometric 
redshifts have been computed with {\sc eazy}, including all available 
UV-optical-NIR bands up to 4.5 $\mu$m \citep[see][]{berta2011}.

We defer to \citet{lutz2011} and \citet{roseboom2010} for further details 
on PACS and SPIRE catalogs, and to \citet{berta2011} and \citet{nguyen2010}
for a derivation of the respective confusion noise. A thorough 
description of photometric redshifts incidence and accuracy in 
the three fields considered is provided in \citet{berta2011}. 
Finally, \citet{berta2011,berta2010} provide additional information about 
ancillary catalogs.

\begin{table}[!ht]
\centering
\begin{tabular}{l c c c}
\hline
\hline
Band	& GOODS-S & GOODS-N & COSMOS \\
\hline
 24 $\mu$m & 0.02 & 0.02 & 0.045 \\  
 70 $\mu$m & 1.1 & --  & --   \\
100 $\mu$m & 1.2 & 3.0 & 5.0  \\ 
160 $\mu$m & 2.4 & 5.7 & 10.2 \\
250 $\mu$m\tablefootmark{a} & 2.6 & 2.3 & 4.8  \\
350 $\mu$m\tablefootmark{a} & 2.2 & 1.9 & 4.0  \\
500 $\mu$m\tablefootmark{a} & 3.1 & 2.7 & 5.7  \\
\hline
\end{tabular}
\tablefoot{\tablefoottext{a}{The SPIRE values refer to instrumental noise and are derived from 
\citet{oliver2012}, using the deepest sections of each field and rescaling to the 3 $\sigma$ 
level. The 3 $\sigma$ confusion noise (after 3 $\sigma$ cut) from \citet{nguyen2010}
 is 11.4, 13.8, 15.6 mJy (at 250, 350 and 500 $\mu$m).}}
\caption{MIPS and \textit{Herschel} 3 $\sigma$ depth $[$mJy$]$ reached in the three selected fields.}
\label{tab:depths}
\end{table}

\begin{figure*}[!ht]
\centering
\includegraphics[width=0.32\textwidth]{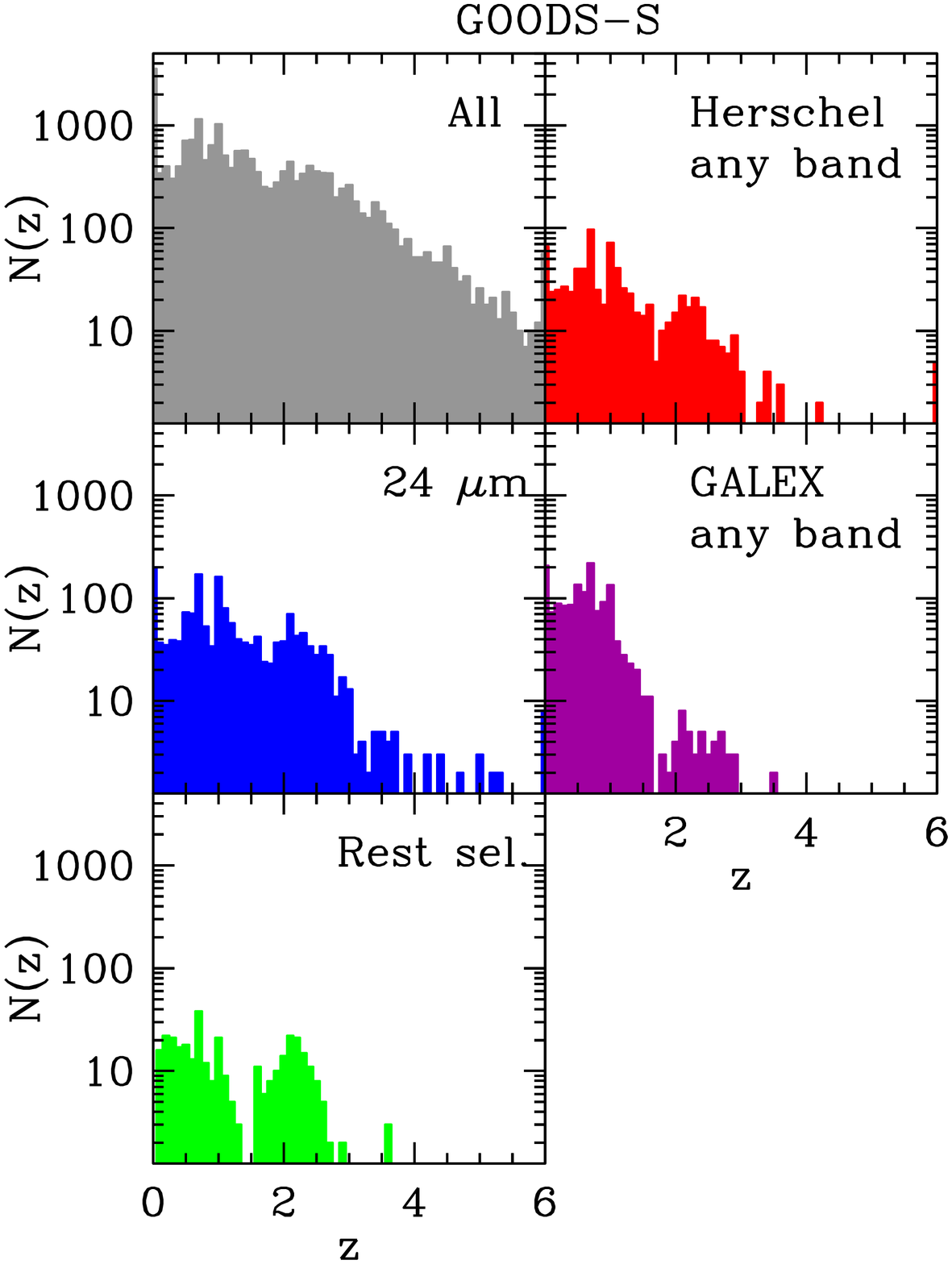}
\includegraphics[width=0.32\textwidth]{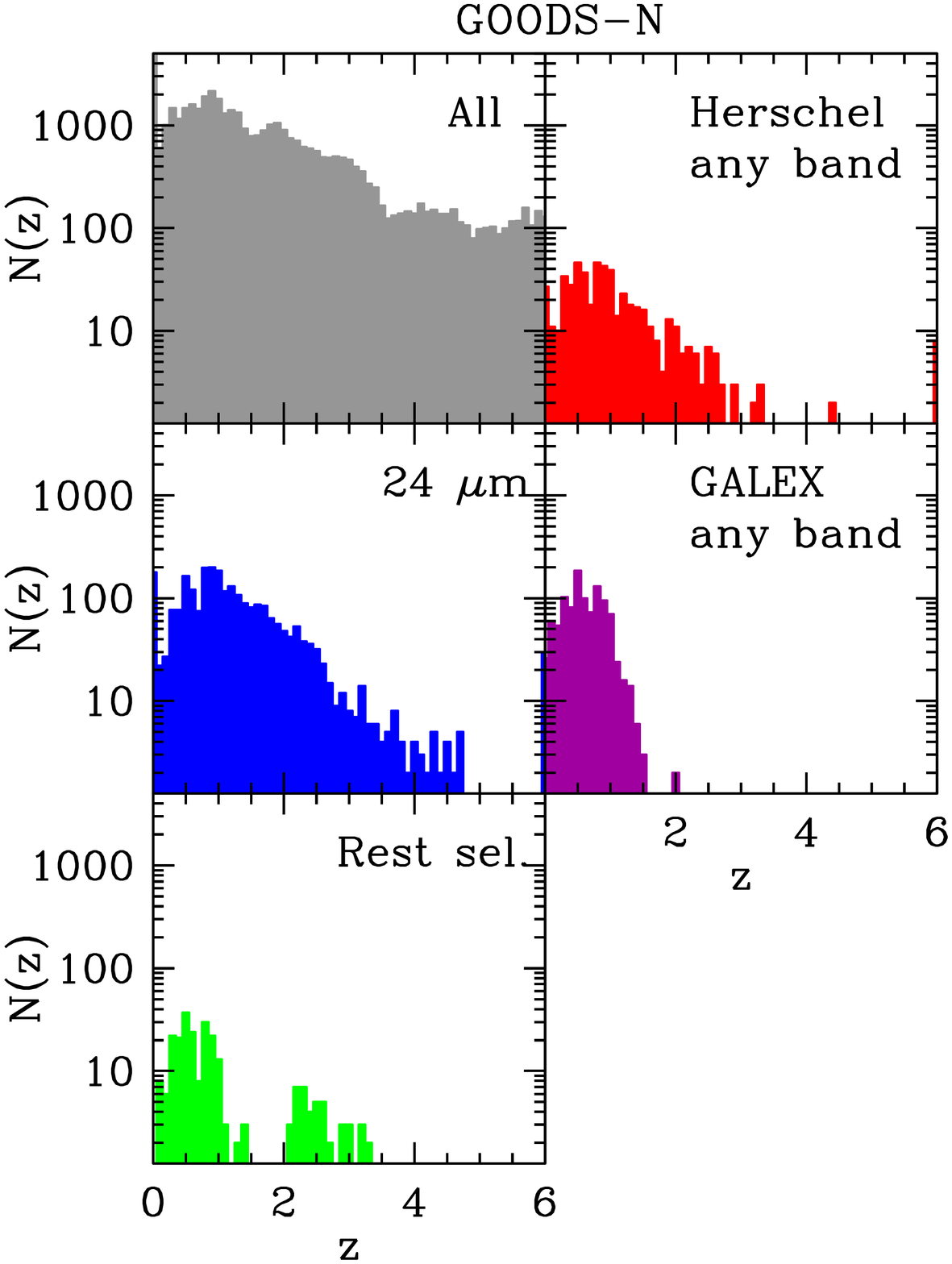}
\includegraphics[width=0.32\textwidth]{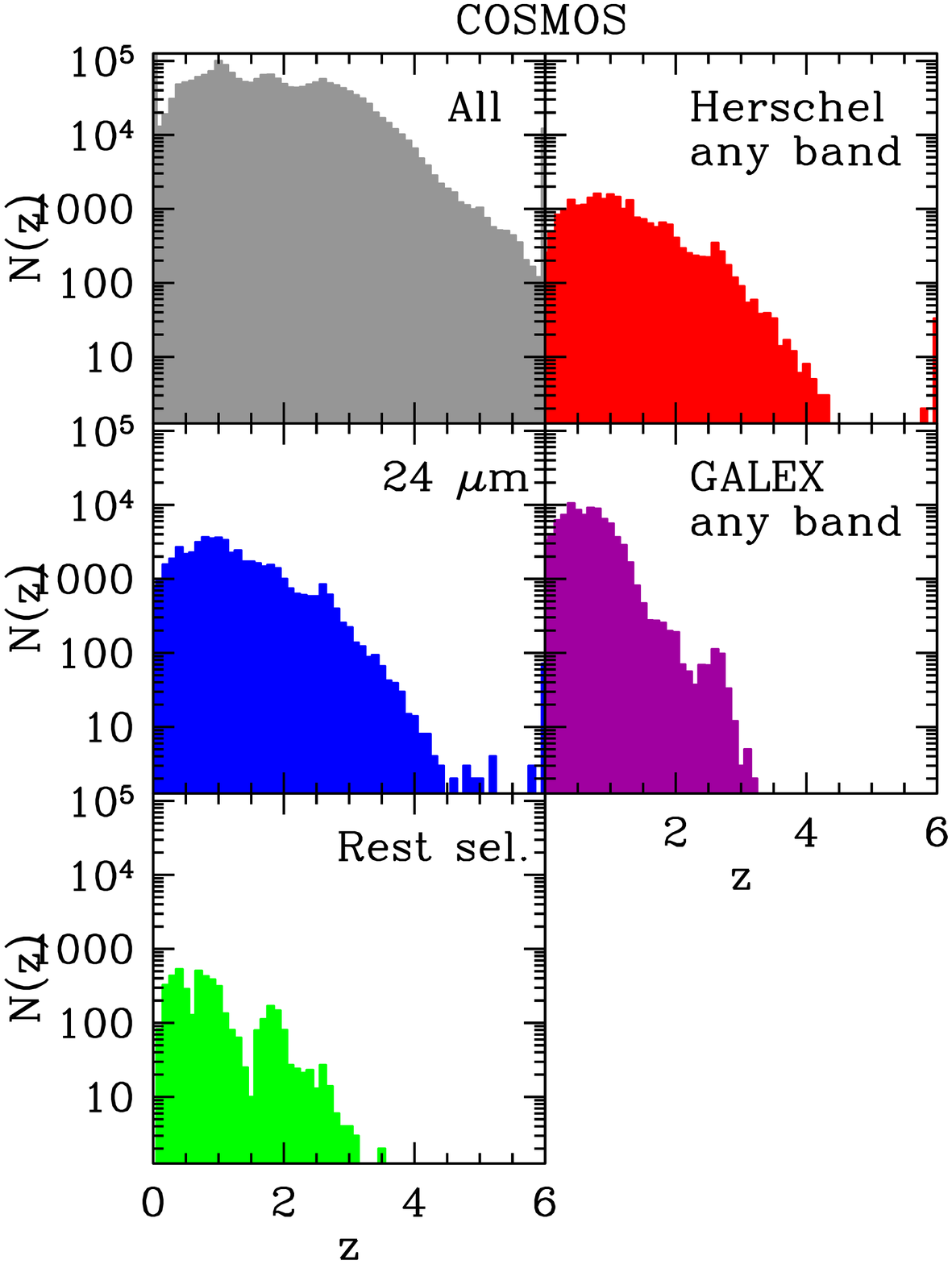}
\caption{Redshift distribution of GOODS-S ({\em left}), GOODS-N ({\em center}) and COSMOS ({\em right}) sources, obtained by applying single-band cuts 
at local $\textnormal{S/N}=3$, and while adopting the restframe interpolation scheme described in the text. For each field, the {\em top-left} panel
 includes all sources in the adopted optical-NIR catalogs, i.e. those built by \citet{grazian2006},
\citet{berta2011}, and \citet{capak2007}.}
\label{fig:selection_effects}
\end{figure*}

\subsection{Restframe colors and source selection}

We are going to group galaxies in an 
$N$-dimensional color space, using a parametric, multi-dimensional technique; 
here we describe the choice of bands and 
colors that will be adopted in this analysis.

Avoiding restrictions in redshift space or galaxy type, we opted to apply the classification technique 
to restframe colors. This approach has the 
advantage of maximizing the number of available sources; moreover 
a classification process in the observed frame would be dominated
by color-redshift dependences.

Adopting a library of SED templates, one could in principle 
derive restframe luminosities, by integrating through the desired passbands, 
but the results would depend on the models/templates adopted.
Alternative methods, such as the one described in \citet{rudnick2003}, 
still imply a hybrid of templates and interpolation, which 
include some dependence on models and the need for a priori SED fitting. 

We hence derive restframe colors by simply 
interpolating between bands -- once redshift is known -- excluding any model assumption. This might 
be a dangerous approximation in some wavelength ranges, where the gap 
between available bands is wide (e.g. in the MIR). 
To avoid this problem, the choice of restframe bands to be used 
is optimized to be fully covered by observed photometry, once 
redshift is taken into account.
Moreover, when performing interpolation, we impose the conditions 
that no more than two adjacent observed bands around each restframe filter
miss a detection.

This approach does not strictly require a detection in each 
band, but only that each restframe filter (in the range between 1400 \AA\ and 100 $\mu$m) is 
bracketed by two observed bands
having detections, with any gap -- due to nondetections -- not larger than two observed bands. 
This allows more sources to take 
part in the grouping, but significantly complicates the selection function. 
Figure \ref{fig:selection_effects} shows the redshift distribution 
of sources in the three fields obtained from \textit{Herschel}, \textit{Spitzer} or GALEX 
brightness cuts and 
resulting from our restframe interpolation scheme. The gap around 
redshift $z\simeq 1.5$ is mainly due to the combination of \textit{Herschel} (or 24 $\mu$m, 
to sample MIR-FIR wavelengths) and GALEX (sampling the UV), with blue optical bands 
covering the ultraviolet regime on the high redshift side.
A total of 234, 343 and 4498 sources from GOODS-N, GOODS-S and COSMOS are 
included, out of 540, 782, and 22659 \textit{Herschel}-detected objects, respectively.

To fulfill the aim of grouping and defining new templates, we need to construct 
a sample representative of the bulk of galaxy populations in color space. 
Nevertheless, because of the requirement to cover as wide a wavelength range as possible, 
the risk of missing some part of color space cannot be fully avoided.
In fact, by imposing a restframe FIR and UV coverage, over a wide range 
of redshifts, we incur into losses of sources mainly due to the limited sensitivity 
of \textit{Herschel} and GALEX surveys. In particular, un-obscured galaxies with 
faint FIR emission might slip away from the selection, and passive galaxies 
are likely poorly represented. In Sect. \ref{sect:outliers} rare objects, outliers of the 
parametric grouping used to define templates, will be identified, thus partially recovering 
parts of color space not sampled otherwise.
Finally, in Sect. \ref{sect:classification} we will compare the new templates -- 
based on the median SEDs of the selected sources -- to observed objects, regardless 
of selection cuts, showing that most of the observed UV-FIR color-redshift space is effectively 
covered.


\section{Parametric description of multi-color data}\label{sect:FastEM}

We describe galaxies by means of their colors, 
as derived from our multi-wavelength catalogs, i.e. their position in
an $N$-dimensional color space. Our aim is to identify structures 
in the $N$-dimensional distribution of galaxies, thus identifying potentially 
different galaxy populations. 

We assume that the $N$-dimensional color distribution function of these sources 
can be reproduced as the superposition of $M$ multivariate Gaussians. 
In what follows, we will refer to these Gaussian components also 
as ``groups'', ``modes'' or ``classes'', without distinction. 
As we are reasoning in restframe, each Gaussian mode would then represent a different
population of galaxies, and can be used to define median colors, to be later fit 
with SED synthesis codes. 

The $M$ components are sought using the code 
by \citet{connolly2000}, further developed and distributed by 
the Auton Lab\footnote{http://autonlab.org/} Team, and now called {\sc FastEM}. The code 
fits the $n$ datapoints, representing our galaxies in the $N$-dimensional space,
with $M$ multivariate Gaussian distributions. {\sc FastEM} uses an expectation-minimization 
algorithm for parameter estimation and a Bayesian information
criterion to decide how many components are 
statistically justified by the data 
\citep{connolly2000,nichol2001a,nichol2001b}, i.e. the number $M$ of 
Gaussians to be used is a free parameter and is optimized 
by {\sc FastEm} on the basis of the actual amount and distribution of 
datapoints.
For each $j$-th Gaussian component, the code 
provides its mean coordinates $\mu_j$ and an $N\times N$ 
covariance matrix $\Sigma_j$ describing its $N$-dimensional shape (including cross terms).

Assuming that the distribution of sources in our $N$-dimensional 
color-space are described by the superposition of 
the $M$ multivariate normal distributions,
the probability of the $i$-th object to belong to
the $j$-th multivariate Gaussian is given by
\begin{equation}\label{eq:probability1}
P\left( x_i , \mu_j , \Sigma_j \right) = A_j \frac{1}{\sqrt{{\left( 2\pi\right)}^N \left|\Sigma_j\right|}}
\exp\left[ -\frac{1}{2}{\left(x_i-\mu_j\right)}^T \Sigma_j^{-1}\left(x_i-\mu_j\right)\right]\textnormal{,}
\end{equation}
where $x_i$ is the position of the given object in the $N$-dimensional space, $\Sigma_j$ is the covariance matrix 
of the $j$-th multivariate Gaussian group, $\mu_j$ is its barycenter (mean coordinate), $A_j$ is its 
amplitude (probability normalization), and the superscript {\small $T$} represent matrix transposition. Note that $x$ and $\mu$ 
are $N$-dimensional arrays and the rank of $\Sigma$ is $N\times N$.

We define as total probability that the $i$-th source belongs to {\em any} of the Gaussian groups, the sum 

\begin{equation}\label{eq:probability2}
P_{\rm tot}\left(x_i\right) = \sum_{j=1}^{M} P\left( x_i , \mu_j , \Sigma_j \right)\textnormal{.}
\end{equation}

This quantity integrates to unity and is therefore a probability density function (PDF),
describing the probability for a single $i$-th object in the sample to lie at the position 
$\mathbf{x_i}=\left(x_{i,1},\ x_{i,2},\ \dots,\ x_{i,N}\right)$ \citep[see][]{davoodi2006}. 

Given Eq. \ref{eq:probability1}, the PDF determines the relative probability 
for any galaxy in the sample to come from any of the $M$ components. We assign each galaxy to 
the mode that maximizes its probability density. \citet{davoodi2006} demonstrated that 
an alternative choice, consisting in randomly assigning each galaxy to one of the modes 
with a probability proportional to the PDF value at its position, gives equivalent results.

\subsection{Grouping in PEP fields}

The three fields adopted for this analysis (GOODS-S, GOODS-N and COSMOS) 
have been observed by PACS and SPIRE down to very different depths (see Table \ref{tab:depths}) 
and over different areas ($\sim 200$ arcmin$^2$ in the GOODS fields and 
$\sim 2$ deg$^2$ in COSMOS). As a consequence, different luminosities 
as a function of redshift are covered, potentially 
probing different regimes with respect to the so-called ``main sequence of star formation'' (MS)
at different redshifts in different fields. 

We therefore chose to keep the three fields separate for classification, in order not to dilute 
the information available. In this way we would like to avoid objects which may be  
peculiar to deep fields (e.g. MS galaxies at high redshift) from getting lost 
in the plethora of COSMOS sources. Furthermore, this choice will lead to a wider variety 
of SED shapes and a better sampling of the observed scatter in the colors.

As a result of the parametric grouping, the color distributions 
of sources in GOODS-S and COSMOS are reproduced by the 
superposition of nine Gaussian modes, while only six components are required 
in GOODS-N. 
The latter misses those groups with bluer MIR-FIR colors.
These are missed also in the large, shallow COSMOS field, where 
-- on the other hand -- a larger population of NIR-MIR power-law sources 
is detected.

Through Monte Carlo simulations, it is possible to estimate the chance that 
one source effectively belonging 
to a given group is nevertheless assigned by {\sc FastEM} to a different one because 
of scatter. 
Based on the groups defined using observed data, we build an 
artificial data-set (see Sect. \ref{sect:simu}), which is then  
evaluated by running {\sc FastEM} using the known set of multivariate Gaussians.
The incidence of mis-classifications -- defined as the fraction 
of sources in a given class being assigned to {\em any}
other group belonging to the same data-set/field -- turns out to vary depending on classes, 
with a maximum of 15\% in the worst cases. Mis-classifications happen 
in the regions of color space where there are overlapping modes
with similar amplitudes.  

Overlap of classes between different fields 
exists. This is evident in subspaces of reduced dimensions, especially due 
to the large dispersion in some colors (see Fig. \ref{fig:sed_fit1}
for error bars on median SEDs),
but becomes difficult to trace when working in the full 10-dimensional set. 
In this context, Table \ref{tab:sed_fits}
lists a subset of colors derived after fitting the median SEDs 
of each Gaussian mode (see next Section). Keeping fields separate results in a finer sampling 
of multi-color space.

\begin{figure*}[!ht]
\centering
\includegraphics[width=0.9\textwidth]{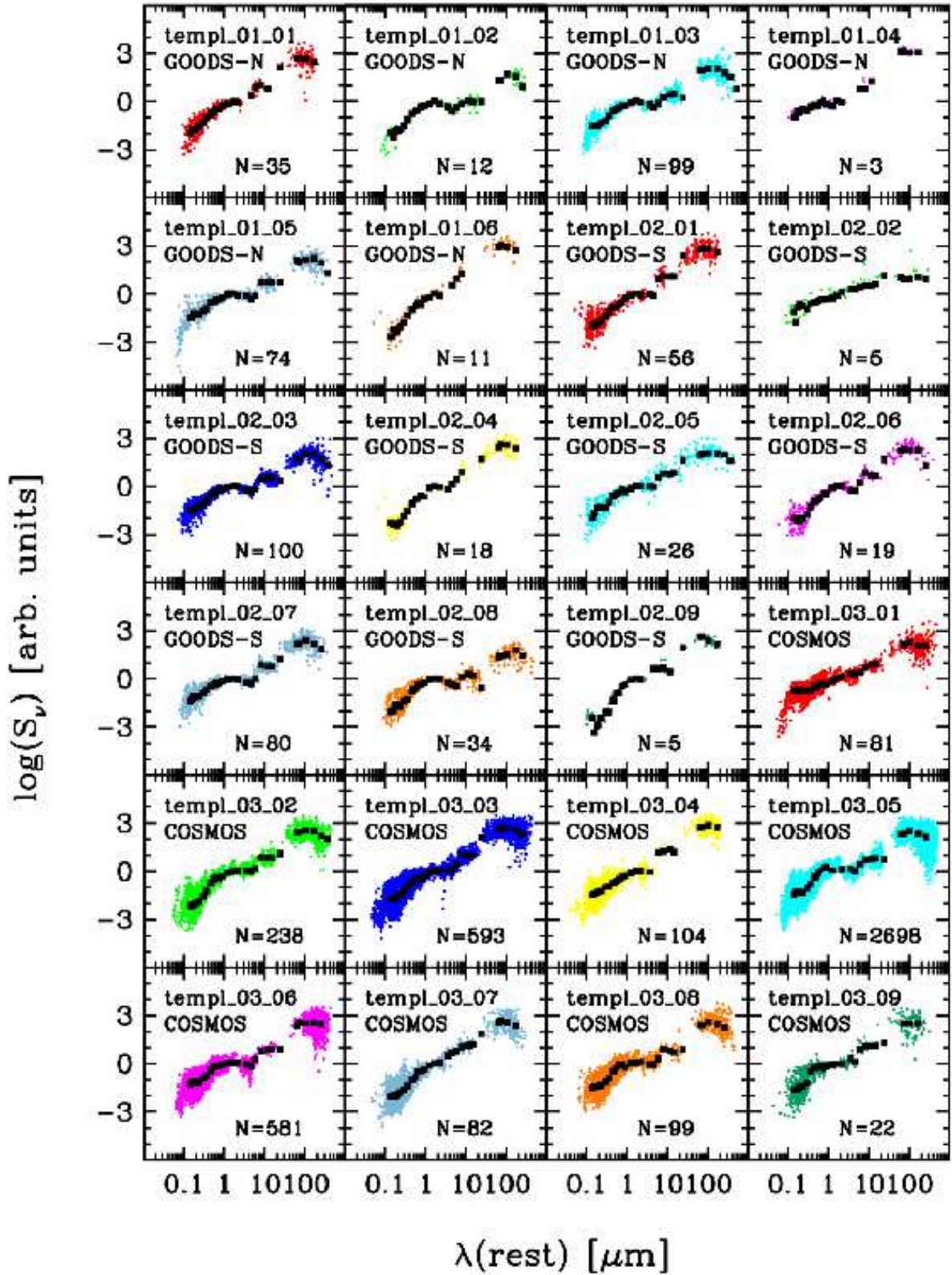}	
\caption{Rest-frame spectral energy distributions of the Gaussian modes found in 
the three fields. Black squares represent median SEDs obtained through restframe filters 
(see Table \ref{tab:avg_sed_filters}). Colored dots trace the de-redshifted 
photometry of individual sources.
The number of objects contributing to each group is 
quoted in the respective panel.}
\label{fig:median_SEDs}
\end{figure*}


\section{Fitting SEDs}\label{sect:sed_fit}

Once that Gaussian modes have been identified, we  
derive the weighted-median\footnote{A weighted percentile 
is defined as the percentage in the total weight is counted, instead of the total number. 
The weighted median corresponds to the 50$^{\rm th}$ 
weighted percentile. In practice, having $N$ values $v_1,\ v_2,\ \dots,\ v_N$ and their corresponding weights
$w_1,\ w_2,\ \dots,\ w_N$, they are first sorted in order of ascending $v$,
and then the partial sum of the weights $S_n=\sum_{k=1}^n w_k$ is computed for each $n$-th (sorted) value.
The weighted median is defined as the value $v_n$ having $S_n=\frac{1}{2} \sum_{k=1}^N w_k$.} 
restframe SEDs of each group. This is done by 
shifting all objects to restframe wavelengths and normalizing at a common 
wavelength: 1.6 $\mu$m in our case. Different normalization wavelengths have 
been attempted, but the NIR turned out to be the most effective choice, as it is well covered by 
the observed bands (i.e. there are no large gaps in wavelength), and it lies in a relatively 
central position of the covered frequency range. Therefore normalizing at 1.6 $\mu$m 
minimizes both uncertainties in interpolation at the normalization wavelength itself and 
scatter at the extremes of the covered spectral range.

\begin{table}[!ht]
\centering
\begin{tabular}{l l}
\hline
\hline
Filter & $\lambda_{eff}$ \\
Name & $[\mu$m$]$\\
\hline
ST-UV14			&	0.1405	\\		
GALEX-FUV		&	0.1539	\\ 	  
ST-UV17			&	0.1751	\\ 	  
OAO-UV2			&	0.1913	\\ 	  
ST-UV27			&	0.22	\\	  
GALEX-NUV		&	0.2316 	\\	  
OAO-UV4			&	0.2986 	\\	  
SDSS-$u$		&	0.3562 	\\	  
ESO-WFI-$B$		&	0.4606 	\\	  
SDSS-$g$		&	0.4719 	\\	  
Subaru-$V$		&	0.5478 	\\	  
SDSS-$r$		&	0.6186 	\\	  
SDSS-$i$		&	0.7506 	\\	  
SDSS-$z$		&	0.8961 	\\	  
ISAAC-$J$		&	1.238  	\\	  
ISAAC-$H$		&	1.652  	\\	  
ISAAC-$K_{\rm s}$ 	&	2.168  	\\	  
IRAC-3.6		&	3.563  	\\	  
IRAC-4.5		&	4.511  	\\	  
IRAC-5.8		&	5.759  	\\	  
IRAC-8.0		&	7.959  	\\	  
IRAS-12			&	11.69  	\\	  
ISO-LW3			&	14.57  	\\	  
MIPS-24			&	23.84  	\\	  
IRAS-60			&	62.22  	\\	  
PACS-070		&	72.48	\\	
PACS-100		&	102.8	\\	
PACS-160		&	165.9	\\	
SPIRE-250		&	251.5	\\	
SPIRE-350		&	352.8	\\	
\hline
\end{tabular}
\caption{Filters adopted to produce rest-frame weighted-median SEDs for each 
Gaussian mode.}
\label{tab:avg_sed_filters}
\end{table}

Median SEDs are computed through a set of passbands, aimed at optimizing
spectral coverage, as listed in Table \ref{tab:avg_sed_filters}.
When computing weighted medians, weights are given by photometric uncertainties 
combined to passbands (i.e. we take the position of de-redshifted points within
the filter transmission curve into account), 
and are propagated into uncertainties on median fluxes by means of an unbiased variance estimator.

Figure \ref{fig:median_SEDs} shows the SEDs of all Gaussian modes in the 
three fields, including de-redshifted photometry of all sources 
belonging to each group (colored points) and median photometry (black squares).

The median SEDs thus obtained are discretized in passbands (Table \ref{tab:avg_sed_filters}), 
and therefore need to be fit with an SED synthesis code to provide a 
reliable ``interpolation'' and description of the whole spectrum.

\subsection{Fitting with {\sc magphys}}

Among the numerous codes available for SED fitting, we adopt the {\sc magphys} 
software \citep{dacunha2008}, because of a number of features:

\begin{itemize}
\item it covers the whole wavelength range from UV to FIR and submm;
\item it links the energy absorbed by dust in the UV-optical domain to dust emission in the MIR and FIR in a self-consistent way;
\item it is user friendly and simple to use; moreover 
the code is structured in such a way that it can be easily modified, if desired.
\end{itemize}

{\sc magphys} combines the Bruzual \& Charlot (\citeyear{bruzual2003}, BC03, see also 
Bruzual \citeyear{bruzual2007}) optical/NIR stellar library, including 
the effects of dust attenuation as prescribed by \citet{charlot2000}, with MIR/FIR dust emission
computed as in \citet{dacunha2008}. The optical and infrared libraries are
linked together, taking into account energy balance (no radiation transfer involved). 
The total energy absorbed by dust in stellar birth clouds and in the ambient 
interstellar medium (ISM) 
is re-distributed at infrared wavelengths. 
The main assumptions are that the energy re-radiated by dust
is equal to that absorbed, and that starlight is the only significant source of
dust heating.

It is worth to note that, since we are fitting the median photometry 
of several galaxies, possible 
variations of the effective attenuation law \citep[e.g.][]{buat2012,reddy2012,reddy2006,wild2011}
are diluted. Such variations would mainly 
influence star formation histories and thus the ages of the 
dominant stellar populations in the {\sc magphys} modeling. Nevertheless, 
our results on infrared luminosities and AGN fractions (see next Sections)
would not be affected by this effect.

We defer to \citet{dacunha2008} for a thorough formal description of 
how galaxy SEDs are built \citep[see also][for an application to \textit{Herschel}-selected 
$z<0.5$ galaxies]{smith2012}. Here we only recall that 
the SED
of the power re-radiated by dust in stellar birth clouds is computed as the sum of three components: 
polycyclic aromatic hydrocarbons (PAHs); a mid-infrared continuum describing 
the emission of hot grains with temperatures $T=130$--$250$ K; and 
grains in thermal equilibrium with $T=30$--$60$ K. 
The ``ambient'' ISM is modeled by fixing the relative proportions of these three 
components to reproduce the cirrus emission of the Milky Way, and adding a component
of cold grains in thermal equilibrium, with adjustable temperature in the range $T=15$--$25$ K.

Different combinations of star formation histories, metallicities and dust content 
can lead to similar amounts of energy absorbed by dust in the stellar birth clouds, and
these energies can be distributed in wavelength using different
combinations of dust parameters. Consequently, in the process of fitting, 
a wide range of optical models is associated with a wide range of infrared spectra and compared to 
observed photometry, seeking for $\chi^2$ minimization. The number 
of possible combinations is on the order of $10^9$ at $z=0$.

Figure \ref{fig:sed_fit1} in Appendix \ref{sect:sed_fit_imgs}
presents the best fits, and Table \ref{tab:sed_fits} summarizes the results.

\subsection{Adding an AGN component}

As mentioned above, one of the main underlying assumptions of the 
{\sc magphys} code is that starlight is the only significant source of
dust heating, i.e. so far a possible AGN component has been ignored 
while fitting our median SEDs. 
We have therefore developed a modified version of the {\sc magphys} code,
adding a warm dust component to the modeled SED emission. This represents 
dust surrounding the active nucleus, often assumed to be distributed 
in a toroidal region (hence hereafter referred as ``torus'' for simplicity).

The \citet{dacunha2008} original code is now combined with the \citet{fritz2006}
AGN torus library \citep[see also][]{feltre2012}. 
The quest for a best fit is still 
based on $\chi^2$ minimization; the stars+dust model is freely normalized 
and subtracted from the observed photometry; 
the torus emission is then added to reproduce what is left 
out from this subtraction. 
Thanks to the {\sc magphys} structure, dust emission --
no more limited to single templates as in some older attempts -- is linked 
to the stellar optical component. 
Allowing the normalization of stars+dust to be free, i.e. not strictly anchored to the observed photometry 
but simply randomly picked from a grid of values, 
the torus is effectively fit to the data in a simultaneous 3-component model 
\citep[see also][for alternative implementations]{fritz2006,berta2007a,noll2009,santini2012,lusso2012,bongiorno2012}.

The torus/AGN library of Fritz et al. spans several geometries of the dust distribution
around the central AGN nucleus, varying the ratio between outer and inner radii ($R_{\rm out}/R_{\rm in}=20$--$300$),
and the aperture angle of the torus (measured starting from the equatorial plane, $\Theta=40^\circ$--$140^\circ$).
The optical depth at the equator covers the range $\tau=0.1-10.0$ at 9.7 $\mu$m. 
The spectrum emitted by the central engine is modeled with a 
broken power-law $\lambda\, L(\lambda)\propto\lambda^\alpha$, with indexes $\alpha_1=1.2$, $\alpha_2=0.0$, $\alpha_3=-0.5$ or $-1.0$,
in the ranges $\lambda_1=0.001$--$0.03$, $\lambda_2=0.03$--$0.125$, and $\lambda_3=0.125$--$20$ $\mu$m, respectively.
Two different sets of models are included, differing for the UV-optical-IR
slope of this power law, $\alpha_3$.
When visible along the line of sight, the direct AGN light is included in these SED models.
The full library comprises a total of roughly 700 models, each one computed at 10 
different lines of sight angles $\Phi$, but we limit the fit to $R_{\rm out}/R_{\rm in}\le100$,
as extreme values have so far not been confirmed by observations \citep[e.g.][]{mullaney2011,hatziminaoglou2008,netzer2007a}.
See \citet{fritz2006} for more details on this model.

In terms of computational time, the performance of our modified code is comparable to 
the original {\sc magphys} code. On an Intel\textsuperscript{\textregistered} 
Xeon$^{\mathrm{TM}}$ 3.2 GHz CPU, $10^{10}$ model combinations 
are sampled in roughly 80--100 minutes, to be compared to 10--15 minutes 
spent for $\sim10^9$ evaluations in {\sc magphys}.

Best fits obtained with {\sc magphys}+AGN are shown in Fig.~\ref{fig:sed_fit1} 
in Appendix~\ref{sect:sed_fit_imgs},
and results are summarized in Table~\ref{tab:sed_fits}.
A description of all physical parameters involved in the fit, and their 
marginalized likelihood distributions (with or without AGN) goes beyond the scope 
of this work. Here we limit to describe the fraction of infrared 
luminosity, $L(8$--$1000\ \mu\textnormal{m})$ powered by the torus component
(see Table \ref{tab:sed_fits}). When sampling the wide parameter space in
searching for a best fit, the code registers the details of all attempted model 
combinations and builds the probability distribution function of a wide number of 
physical quantities. In this way degeneracies in the fit are taken into 
account and are translated into uncertainty ranges. Table \ref{tab:sed_fits}
includes the AGN fraction of the best fit, as well as its 2.5$^{\rm th}$, 50$^{\rm th}$ and 97.5$^{\rm th}$ 
percentiles, and gives a short description in terms of their main properties
(AGN component, star formation strength, colors, etc).


\section{Rare sources}\label{sect:outliers}

The probability of a given object to belong to
any of the Gaussian modes found by {\sc FastEM} and 
its total probability to belong to {\em any} of these groups is described by 
Eqs. \ref{eq:probability1} and \ref{eq:probability2}.
We build the cumulative total probability distribution function over all 
objects in each field and perform Monte Carlo simulations in order to identify
a potential excess of observed objects at the low probability end. 

\begin{figure}[!ht]
\centering
\includegraphics[width=0.35\textwidth]{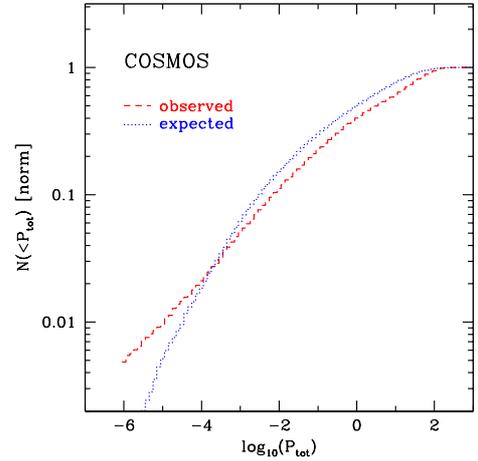}
\caption{Cumulative distribution of probability densities in COSMOS. The red dashed line
is for the observed sources, and is compared to simulated results (blue dotted line).
The excess of observed sources, with respect to expectations from simulations, at the low 
probability side is due to rare objects, not complying with the restframe colors of 
Gaussian groups.}
\label{fig:total_pdf_cos}
\end{figure}

Such an excess can be used to define potential {\em outliers}, or rare sources 
which cannot be categorized in any of the Gaussian modes found. 
Figure \ref{fig:total_pdf_cos} shows the cumulative $P_{\rm tot}$ distribution 
in COSMOS (red dashed line).

\begin{figure*}[!ht]
\centering
\includegraphics[width=0.75\textwidth]{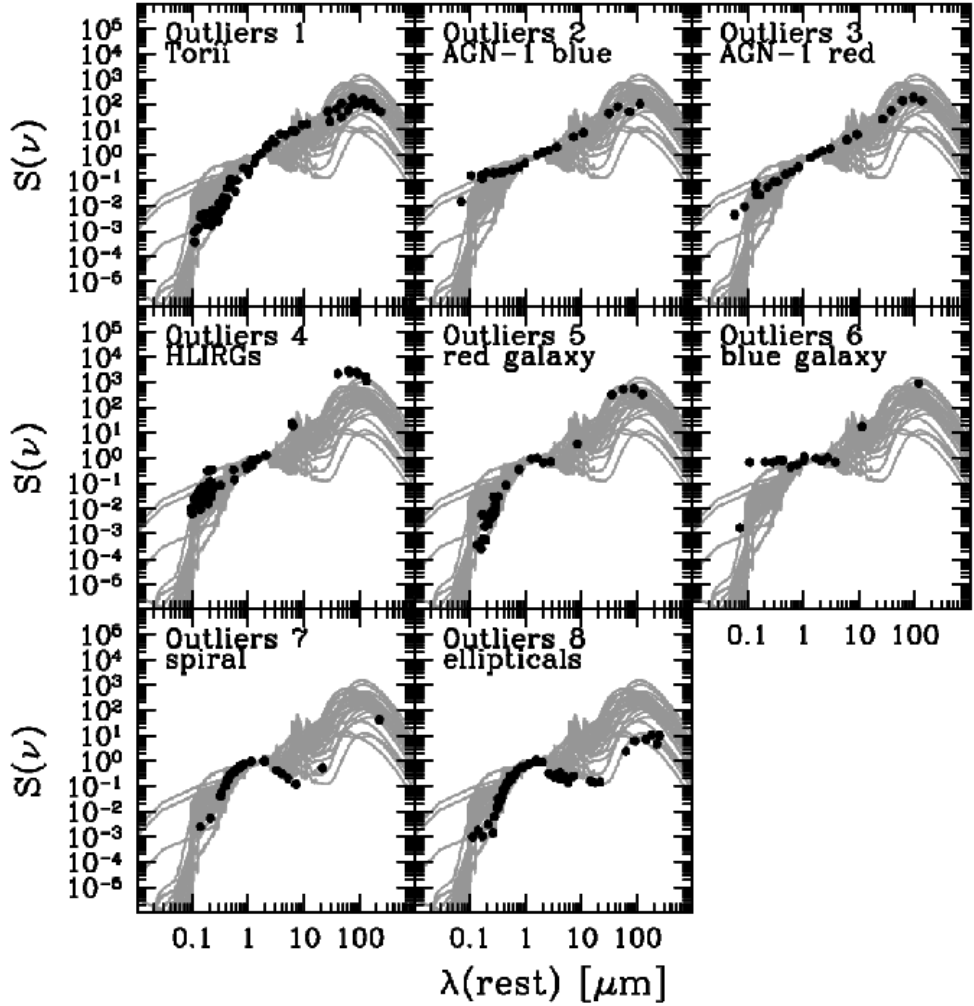}
\caption{SEDs of outliers (black symbols), grouped in classes, and compared to the 
rest of {\sc FastEM}-based templates (grey), normalized to 1.6 $\mu$m. }
\label{fig:outlier_SEDs}
\end{figure*}

\subsection{Simulations}\label{sect:simu}

Given the $M$ multivariate normal distributions that build up our $N$-dimensional color-space, 
we run a Monte Carlo simulation aimed at constructing the expected total probability distribution 
function PDF$_{\rm exp}$. The code designed for this goal, produces a set of $n$
random points for each $j$-th $N$-dimensional multivariate normal distribution, 
where $n$ is the actual number of real sources belonging to the $j$-th group.
Then the artificial catalogs thus built are ``evaluated'' by {\sc FastEM} 
using the multivariate normal groups previously found on real catalogs, and 
already used to build the random artificial catalogs themselves.
The probability of each artificial object to belong to any of the $j=1,\, \dots,\, M$ multivariate 
distributions (i.e. including all others that were {\em not} used to build the random 
sample) is thus computed in the same way as for real objects (Eq. \ref{eq:probability2}). 
The $M$ 
probabilities are summed together for each of the $i=1,\, \dots, n\times M$ random points in
the $N$-dimensional space. The expected cumulative distribution of 
probability densities expected from simulations (using all random 
points) is then compared to the ``observed'' one in Fig. \ref{fig:total_pdf_cos} (blue dotted line).

The analysis of these simulations is not straightforward. In principle, we 
would like to define a probability threshold $P_{\rm thresh}$, below which a source can 
be considered an {\em outlier} \citep[see][]{davoodi2006}.
In practice, in the COSMOS case, the observed and simulated curves intersect each other
at roughly $P_{\rm tot}\simeq10^{-4}$. In our case $N(\textnormal{obs})=2\times N(\textnormal{exp})$ 
holds at $P_{\rm tot}\simeq10^{-5}$. 
Simulations for the GOODS fields (not shown here for simplicity) 
have a different behavior:
the observed and simulated PDFs do not cross each other, but are 
roughly parallel, with PDF$_{\rm exp}\gtrsim$PDF$_{\rm tot}$.
The two GOODS fields include a rather limited number of 
sources, i.e. our $N$-dimensional space is rather sparsely sampled. In such a 
small-number limit, simulated and real probabilities
follow the same trend, but real catalogs are missing a fraction of sources. 
On the other hand, the large COSMOS (Fig. \ref{fig:total_pdf_cos}) indeed contains 
some significant fraction of unexpected objects at the low-probability end. 
To be conservative, we adopt $P_{\rm thresh}=10^{-5}$ as threshold for 
all fields.

\subsection{Outliers}

Using the probability threshold defined above, we identify outlier candidates 
in the $N$-dimensional color space. In this way, 5, 21, and 32 potential outliers are 
identified in GOODS-N, GOODS-S and COSMOS, respectively.

The SEDs of selected objects are checked 
individually against photometric oddities (e.g. glitches, cosmic ray hits, 
etc.) and multi-wavelength images (in the F814W or F850LP ACS bands, together with 3.6 $\mu$m, 24 $\mu$m, 100 $\mu$m, 250 $\mu$m, and 
others if needed) are visually inspected. 
Several outlier candidates turn out to be sources affected by potential
photometric problems in some band, mostly 
due to blending, wrong associations, and (consequently) also wrong photometric redshifts. 
These cases are often related to each other. 
For example a faint galaxy in the vicinity of a bright star would have an optical SED
resembling that of a passive object, but would be also bright in the FIR. 
The warning bell in such a case would be rung by the fact that 
the 1.6 $\mu$m stellar peak would appear to lie at the wrong wavelength, 
with respect to the tabulated redshift. Another exemplary case is an object appearing isolated 
in the optical, but double at 24 $\mu$m and finally being the blend of two components 
in \textit{Herschel} maps.

In addition, objects at the edge of a given {\sc FastEM} group definition 
might be identified as outlier candidates, but it might happen that 
they are reproduced by the SED of a neighboring mode. 
Since we are here searching for objects not covered by any of the SEDs defined so far, 
these cases are not considered as real outliers. 

After these individual checks, 12 definite outliers are left. Their SEDs are shown in 
Fig. \ref{fig:outlier_SEDs}, and compared to the median SEDs of {\sc FastEM} groups
(see Sects. \ref{sect:FastEM} and \ref{sect:sed_fit}). We have assigned them to eight new 
groups, according to their SEDs and visual inspection of ACS images.

Outliers of the {\bf group 1} are two GOODS-S and one COSMOS sources with torus-dominated, almost featureless SEDs. 
They are characterized by moderate FIR emission, a prominent MIR excess and red optical 
colors. All three are known X-ray sources \citep{luo2008,luo2010,brusa2010,cappelluti2009} 
and two are detected at radio frequencies \citep{miller2008,schinnerer2010}.

Galaxies in {\bf groups 2 and 3} have featureless continua from the UV to the MIR and their 
IRAC-MIPS emission is likely dominated by an AGN component. The two differ from each other 
in their optical colors. Both SEDs have flat power-law like shape, while group 1 is convex
in the NIR-MIR regime. Both are detected in the X-rays \citep{luo2008}.
The prototype galaxy of group 3 is a bright radio emitter with a 1.4 GHz
flux of $\sim3.31\pm0.01$ mJy at $z\simeq1.6$, while the other is not detected at radio frequencies \citep{miller2008}.

{\bf Group 4} includes two $z\sim3$ galaxies with very bright FIR and far- to mid-infrared flux
ratio much redder than any class found by {\sc FastEM}. 

Outliers of {\bf class 5} represent bright FIR emitters with very red optical 
SEDs. With an observed flux ratio $S_\nu(100)/S_\nu(24)=90$ and a redshift $z\sim1.8$, 
this outlier is similar to the high redshift ``silicate break'' 
galaxies described by \citet{magdis2011}, although no 16 $\mu$m information 
is available in our case.

Outliers of {\bf class 6} are represented by a $z\sim1.1$ Lyman break galaxy, 
detected well in the near-UV band by GALEX, but barely seen in the far-UV. The optical emission is very blue and 
does not comply with any previously defined SED. 

{\bf Class 7} includes a nice and isolated spiral galaxy at $z=0.11$, with no 
MIR emission and very weak FIR.

Finally, {\bf group 8} comprises two large nearby spheroidal galaxies at redshifts $z=0.10$ and $z=0.38$.
They both are characterized by a slight MIR excess, with 
respect to a pure old stellar population Rayleigh-Jeans spectrum and faint FIR emission, 
which can be interpreted as circumstellar dust of AGB stars and emission 
from diffuse dust even in these passive systems \citep[see, for example,][]{bressan2006,panuzzo2011}.

As in the case of the Gaussian modes, the SEDs of outliers have been 
fit using the original {\sc magphys} \citep{dacunha2008} and our modified version including 
an AGN component. Results are shown in Appendix \ref{sect:sed_fit_imgs} and 
summarized in Table \ref{tab:sed_fits}.


\section{Discussion}\label{sect:discussion}

Combining the modeled SEDs of multivariate Gaussian modes and outliers, 
a library of new templates spanning from the 
ultraviolet to submillimiter spectral domains is now defined. 

Here we apply this new library to full multi-wavelength 
data sets in GOODS-N and GOODS-S, with the aim of testing its performance 
in evaluating infrared luminosities of galaxies and 
classifying 24 $\mu$m-selected objects on the basis of their 
AGN content, FIR colors and far-to-near infrared flux ratios.

\subsection{Deriving infrared luminosities}\label{sect:testing}

We would now like to compare the new library of templates to other existing 
templates and methods, focusing our attention 
on the estimate of infrared (8--1000 $\mu$m) luminosity. 

Figure \ref{fig:all_templates} shows all templates, as in Table \ref{tab:sed_fits},
normalized to 1.6 $\mu$m and color coded by their $L(8$--$1000\, \mu\textrm{m})/L(1.6\, \mu\textrm{m})$ 
color. 
A normalization in correspondence of the NIR stellar emission peak stresses 
color differences and highlights the wide variety of FIR SEDs describing real 
sources.

The PEP team defined a test case catalog, extracted semi-randomly from 
the GOODS-S MUSIC+PEP source list \citep{lutz2011,berta2010,berta2011,grazian2006}, 
aimed at comparing different $L(\textnormal{IR})$ estimators. 
The list of sources was designed to cover as wide a redshift and PACS flux
range as possible, and includes 200 entries. No SPIRE data were included at this stage.
In this analysis, the following approaches are being tested:

\begin{figure}[!ht]
\centering
\includegraphics[width=0.45\textwidth]{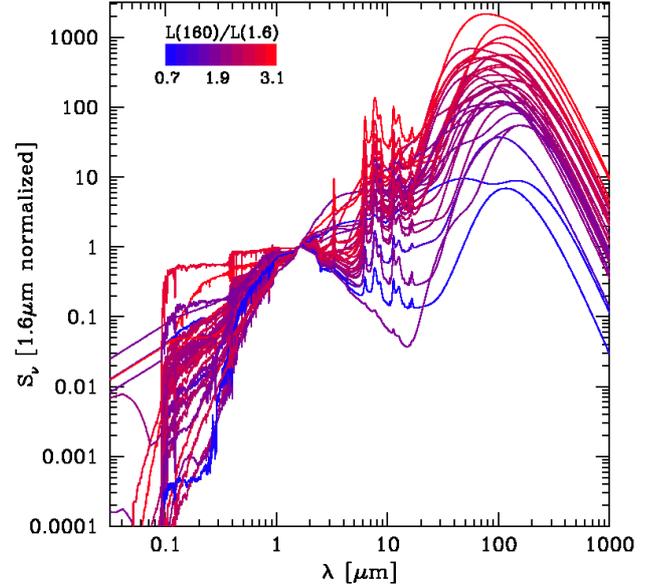}
\caption{SEDs of all templates, belonging to all groups and outliers.
Color coding is based on the infrared bolometric 
(8--1000 $\mu$m) to NIR (1.6 $\mu$m) luminosity ratio.}
\label{fig:all_templates}
\end{figure}

\begin{figure*}[!ht]
\centering
\includegraphics[width=0.9\textwidth]{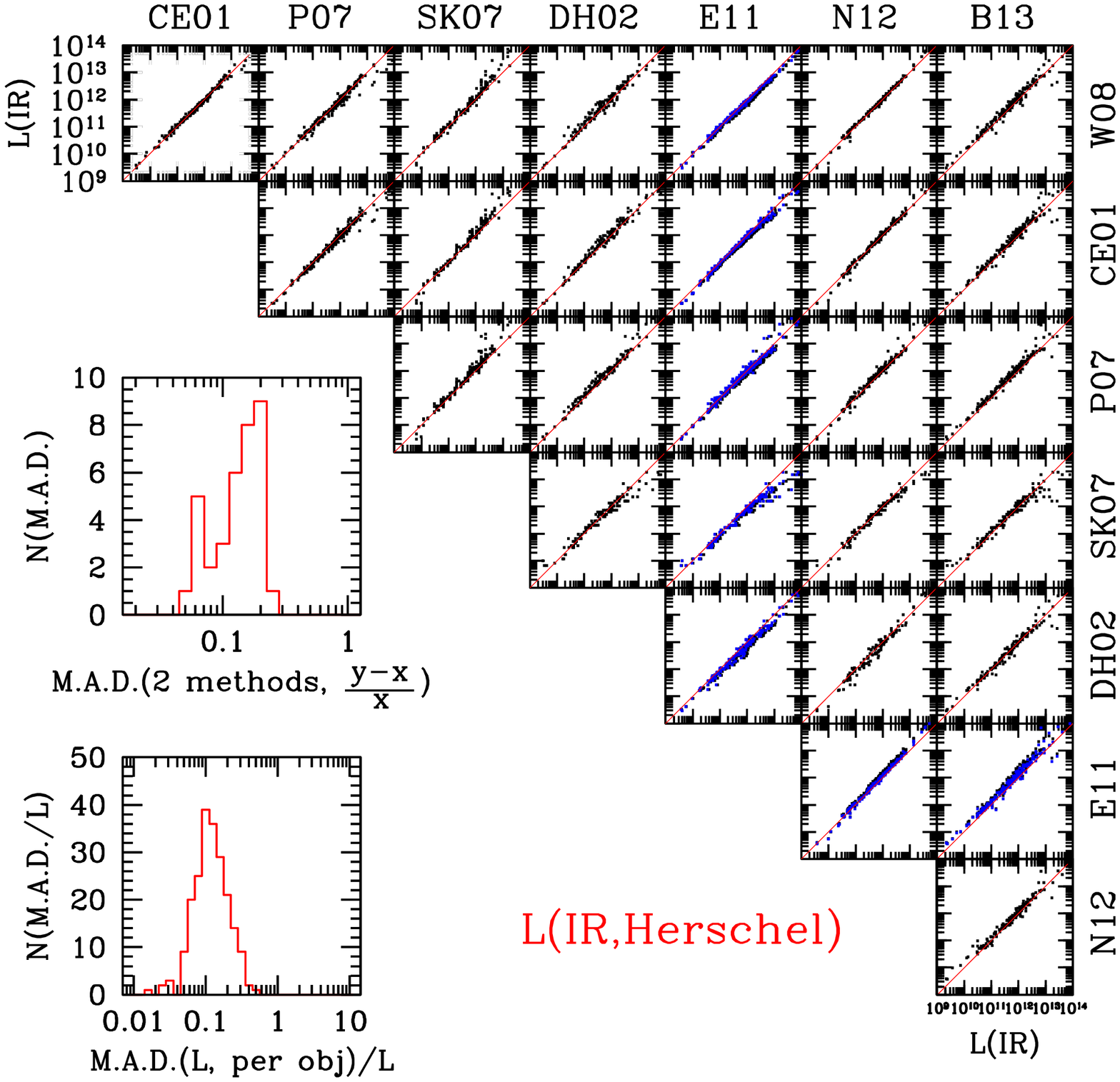}
\caption{Testing $L(\textnormal{IR})$ estimates using 8 different methods, and including 
\textit{Herschel} data. The employed methods and templates are: \citet[][W08]{wuyts2008}; 
\citet[][CE01]{chary2001}; \citet[][P07]{polletta2007} fitted using {\sc hyper-z}; 
\citet[][SK07]{siebenmorgen2007} fitted with a custom code; \citet[][DH02]{dale2002};
\citet[][E11]{elbaz2011}; \citet[][N12]{nordon2012}; and this work (B13). 
\citet{elbaz2011} adopt two different templates (``main sequence'', shown in black, and ``starburst'', in blue).
See text for more detail about each method. Red diagonal lines simply mark the one to one 
relation.
The large bottom left panel shows the distribution of the relative spread per object, 
computed as $\textnormal{M.A.D.}/\textnormal{median}$ using the 8 different $L(\textnormal{IR})$ 
estimates per object. The large upper left panel, instead, depicts the distribution of
scatter computed for each pair of methods as M.A.D. of the quantity $\left(y-x\right)/x$. 
In this case, $x$ and $y$ denote pairs of $L(\textnormal{IR})$ estimates on the 
abscissa and ordinate axes, for each pair of methods.}
\label{fig:test_LIR_1}
\end{figure*}

\begin{figure*}[!ht]
\centering
\includegraphics[width=0.6\textwidth]{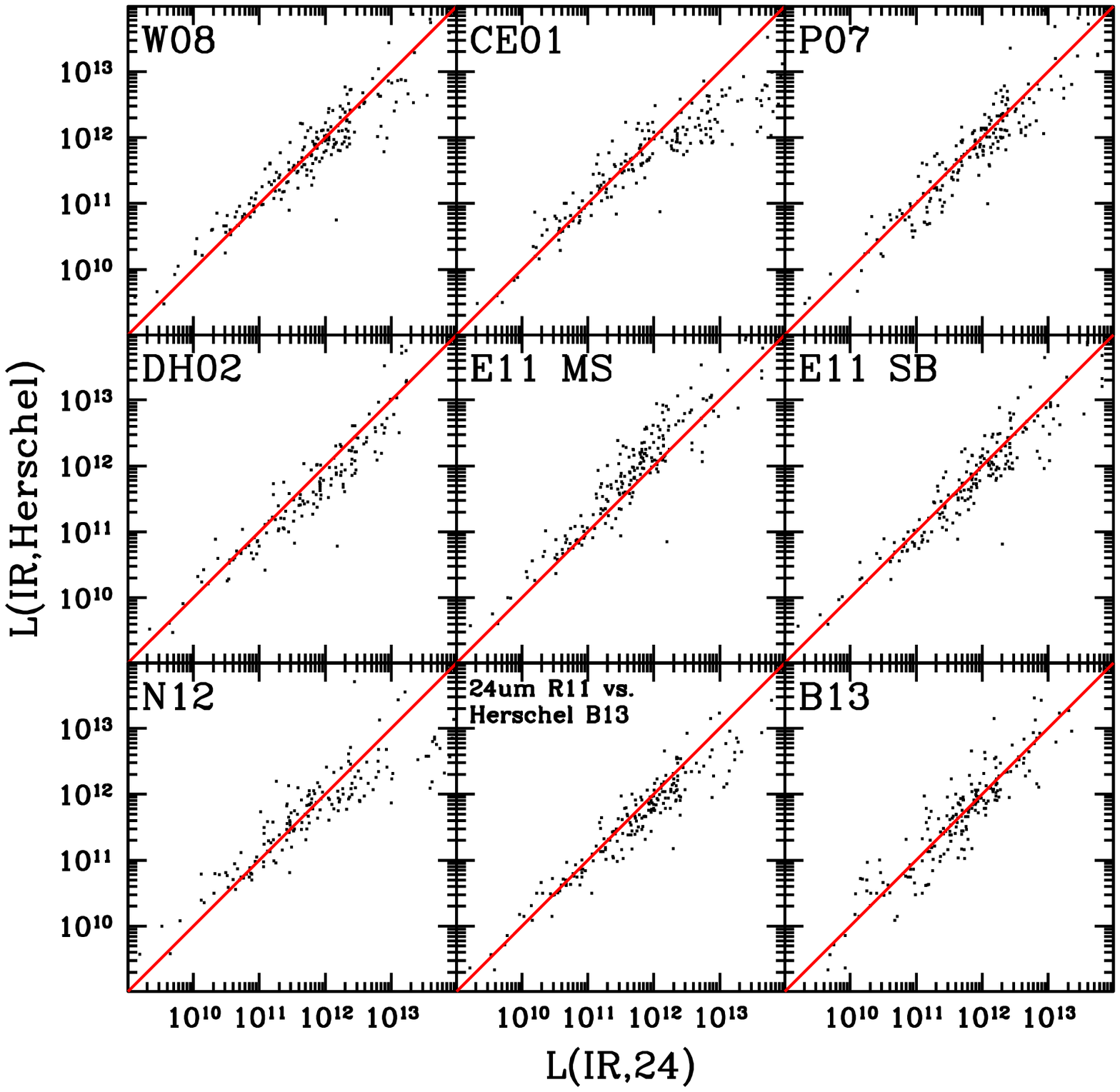}
\caption{Comparison of infrared luminosities, as obtained with and without \textit{Herschel} data 
included in the fit. In the latter case, 24 $\mu$m flux densities are used. 
The methods and templates used are: \citet[][W08]{wuyts2008}; 
\citet[][CE01]{chary2001}; \citet[][P07]{polletta2007} fitted using {\sc hyper-z}; 
\citet[][SK07]{siebenmorgen2007} fitted with a custom code; \citet[][DH02]{dale2002};
\citet[][E11]{elbaz2011}; \citet[][N12]{nordon2012}; \citet[][R11]{rujopakarn2011}; and this work (B13).
The fit obtained with SK07 templates was designed to 
work with multi-band photometry, hence was not tested on 24 $\mu$m data alone and is not shown here.
\citet{rujopakarn2011} provide an estimate of $L(\textnormal{TIR})$ using 24 $\mu$m only, but no \textit{Herschel} information, therefore 
it is here compared to the results obtained with the new templates (B13) including \textit{Herschel} data. 
A correction to transform $L(\textnormal{TIR}, 5-1000\, \mu\textnormal{m})$ 
into $L(\textnormal{IR},8$--$1000\, \mu\textnormal{m})$
has been computed by integrating \citet{rieke2009} templates, as used in \citet{rujopakarn2011}; 
this correction ranges between 8\% and 1\% depending on luminosity. 
Red diagonal lines simply mark the one to one relation.}
\label{fig:test_LIR_2}
\end{figure*}

\begin{enumerate}
\item the luminosity-independent conversion between observed flux density and
luminosity by \citet[][W08]{wuyts2008}, based on a single template,
constructed by averaging the logarithm of \citet[][DH02]{dale2002} templates.
In terms of local analogs, its mid- to far-infrared SED shape is reminiscent of M82.
\item fit of FIR SEDs with the \citet[][CE01]{chary2001} luminosity-dependent 
template library \citep[see also][]{nordon2010,elbaz2010,hwang2010}.
\item standard UV-to-FIR SED fitting using the {\sc hyper-z} code \citep{bolzonella2000} and 
\citet[][P07]{polletta2007} templates, allowing for extinction with $A_V=0$--$3$.
\item maximum likelihood analysis using a custom code and the starburst SED library 
by \citet[][SK07]{siebenmorgen2007}, which is based on radiative transfer and spans
a 3 dimensional parameter grid of luminosity, nuclear radius, and visual
extinction.
\item fit of PACS flux densities using DH02 templates, normalized 
to an IR luminosity (8--1000 $\mu$m) using the \citet{marcillac2006} relation 
between $L(\textnormal{IR})$ and the $F(60)/F(100)$ color, and corrected on 
the basis of \citet{papovich2007} recipe to account for the overestimation
of 24\,$\mu$m-based $L(\textrm{IR})$ \citep[see][]{santini2009}.
\item rescaling of the two \citet[][E11]{elbaz2011} templates of ``main sequence'' (MS) and ``starburst'' (SB)
galaxies to observed FIR data. 
\item the \citet[][N12]{nordon2012} recipe, based on the offset from the star-forming ``main sequence'', 
$\Delta(\textnormal{sSFR})_{\rm MS}$, to re-calibrate CE01 
templates. 
\item fit of UV-FIR data, using {\sc hyper-z} and the templates defined in this work (hereafter labeled B13). 
\end{enumerate}

In all cases, redshifts have been fixed to known values 
(see Sect. \ref{sect:data} for details). 
Estimations of $L(\textnormal{IR})$ with each method have been performed both including and 
excluding \textit{Herschel} data. In the latter case, 24 $\mu$m data were used in the  
estimation of infrared luminosities. Most approaches are based on $\chi^2$ minimization
while fitting the data. Methods number 6 and 7 need additional knowledge of stellar mass 
(and SFR). Ideally, method 6 would also require 
an estimate of either IR8 (i.e. $L_{\rm IR}/L_{8 \mu{\rm m}}$) or 
compactness, in order to distinguish between the MS and SB templates. Since these
pieces of information are not available in our case, we apply both templates to all objects, 
thus deriving two $L(\textnormal{IR})$ values for each source.

In addition to the above listed techniques, when excluding \textit{Herschel} data 
we also applied to our test case the recipe 
by \citet[][R11]{rujopakarn2011}, mapping 24 $\mu$m fluxes into total 
infrared luminosities $L(\textnormal{TIR})$. This method was defined by
parameterizing galaxy SEDs as a function of luminosity surface density, 
$\Sigma_{L({\rm TIR})}$, and adopting the \citet{rieke2009} templates.
The correction needed to transform $L(\textnormal{TIR}, 5-1000\, \mu\textnormal{m})$ 
into $L(\textnormal{IR},8$--$1000\, \mu\textnormal{m})$
has been computed by integrating these templates; 
this correction ranges between 8\% and 1\%, depending on luminosity.

Figure \ref{fig:test_LIR_1} shows the direct comparison of $L(\textnormal{IR})$ 
estimates based on the different techniques. 
Infrared luminosities derived including \textit{Herschel} data are 
globally very consistent with each other. When computing the 
dispersion of all 8 determinations around their median value for each 
object (i.e. computing the median absolute deviation, M.A.D., per object), it turns 
out that the relative uncertainties peak at $\sim$10\% with a tail 
extending to $>50$\% for a few cases only (large bottom-left panel in Fig. \ref{fig:test_LIR_1}). 
In other words, by implementing \textit{Herschel} photometry in SED fitting, 
it is possible to estimate $L(\textnormal{IR})$
within $\sim10$\% (M.A.D.), regardless of the adopted technique.
No trends of M.A.D. as a function of redshift or infrared luminosity are 
detected. 

To quantify the scatter in each possible combination of two methods
shown in Fig. \ref{fig:test_LIR_1}, the upper large panel on the left presents 
the distribution of the M.A.D. of the quantity $\left(y-x\right)/x$, computed for 
each pair of methods using all sources. In this case $x$ and $y$ simply 
represent the $L(\textnormal{IR})$ values computed with methods in the abscissa or 
ordinate axis. When comparing two methods, the scatter over all objects 
ranges between a few percent to $\sim20$\% (in M.A.D. terms again). 
Most of the adopted 
techniques present only few catastrophic outliers, defined as those sources
having $\left| \frac{y-x}{x} \right|>2.0$, when compared to other methods.

In some cases the scatter between two methods is remarkably small, 
pointing to their similarities. 
The \citet{chary2001}, \citet{elbaz2011} and \citet{nordon2012} approaches are 
conceptually very similar and produce the lowest scatter when compared to 
each other. Also the \citet{wuyts2008} approach produces scatter lower than 10\%,
when compared to the above mentioned three techniques. 
These four $L(\textnormal{IR})$ estimates practically reduce to  
fitting a single template to the observed photometry, either chosen 
from a library on the basis of luminosity itself or the offset from the ``main sequence''
 of star formation $\Delta(\textnormal{sSFR})_{\rm MS}$ (CE01, N12), or 
simply being the only available choice (W08, E11).

As the complexity of fits increases, the scatter also grows: techniques employing 
a variety of templates with free normalization, and/or extending 
the wavelength range all the way to the NIR, optical or UV, naturally produce 
a larger scatter, when compared to others.
It is worth pointing out (not shown for the sake of conciseness) that, in 		   
UV-to-FIR SED fitting,
scatter seems to be dominated more by the choice of code and its setup, rather than specific templates. 
Fitting SEDs with the new template library or with Polletta's, 
but adopting {\sc hyper-z} \citep{bolzonella2000}, {\sc eazy} \citep{brammer2008}, 
or {\sc Le Phare} \citep{arnouts1999,ilbert2006} produces significantly 
different amounts of scatter. Determining the source 
of this effect is not trivial, because tuning these
codes so that the adopted configurations are equivalent is not 
straightforward. 
To give an idea of the size of this effect, when considering these three codes 
with standard setup  
and employing the same template library, the peak of the distribution of relative 
M.A.D. scatter per source (see the large bottom-left panel in Fig. \ref{fig:test_LIR_1} for comparison) can 
shift to values as high as 20\%.

Deep \textit{Spitzer} blank field observations at 24 $\mu$m contain a large number of objects 
not detected by \textit{Herschel}.
It is thus important to verify how reliable the 
$L(\textnormal{IR})$ derivation will be in the absence of FIR photometry. 
This is particularly relevant in light of the short \textit{Herschel} lifetime and because of the spectral range
covered by other 
current/future space missions (e.g. WISE, 3--25 $\mu$m; JWST, 
0.6--28 $\mu$m), before the launch of next generation projects extending 
to the FIR again (e.g. SPICA).

Figure \ref{fig:test_LIR_2} compares $L(\textnormal{IR})$ estimates obtained with 
and without the inclusion of \textit{Herschel} data, for each approach.
The well known bending in luminosity-dependent, locally-calibrated (e.g. CE01-based) approaches 
\citep[see][]{elbaz2010,nordon2010} 
and its correction \citep{elbaz2011,nordon2012} can be seen here. 
It is worth noting that -- since higher luminosity objects tend to lie at a higher redshift because of Malmquist bias --
the deviation from the one to one line affects mainly 
distant galaxies \citep[$z\ge1.5$, see also][]{elbaz2010,elbaz2011}, while all methods 
show self consistency below $L(\textnormal{IR})\simeq10^{12}$ L$_\odot$, i.e. at lower redshift.
We warn that since the knowledge of IR8 (i.e. $L_{\rm IR}/L_{8 \mu{\rm m}}$) or 
compactness is not available, a proper choice between the \citet{elbaz2011} MS or SB 
templates is not possible; thus a direct comparison 
between the two should not necessarily be taken at face value for this test case.
Classical SED-fitting methods, allowing for template luminosity rescaling, 
seem to be less affected by this problem at the high luminosity end, highlighting conceptual 
differences between linear and nonlinear $L(\textnormal{IR})$ derivations/corrections.
\citet{nordon2012} analyze in detail the systematics in $L(\textnormal{IR},160)/L(\textnormal{IR},24)$
affecting luminosity-dependent methods at $z\sim2$, showing that N12 seems to be the 
most effective one in reducing trends, at the expense of a 1.2--1.5 times larger scatter
with respect to other approaches. For comparison, when extending to all redshifts (Fig. \ref{fig:test_LIR_2} here), the scatter 
in $L(\textnormal{IR,\textit{Herschel}})/L(\textnormal{IR},24)$ is $\sim0.3$ dex (in terms of $\sigma$)
for most methods, with the exception of CE01 and N12 ($\sim0.45$ dex).

Summarizing, when including \textit{Herschel} photometry,
all libraries and methods provide $L(\textnormal{IR})$ estimates
within a $\sim10-20$\% scatter, modulo tuning subtleties;
known caveats \citep[see, e.g.,][]{elbaz2010,nordon2010} apply to 
derivations limited to MIR wavelengths. The latter 
are still affected by a large scatter, when compared to FIR-based 
$L(\textnormal{IR})$ estimates, thus pointing out the importance of 
\textit{Herschel}-based surveys and future upcoming FIR missions.


\subsection{Classification of MIR and FIR sources}\label{sect:classification}

\begin{figure*}[!ht]
\centering
\includegraphics[height=0.90\textheight]{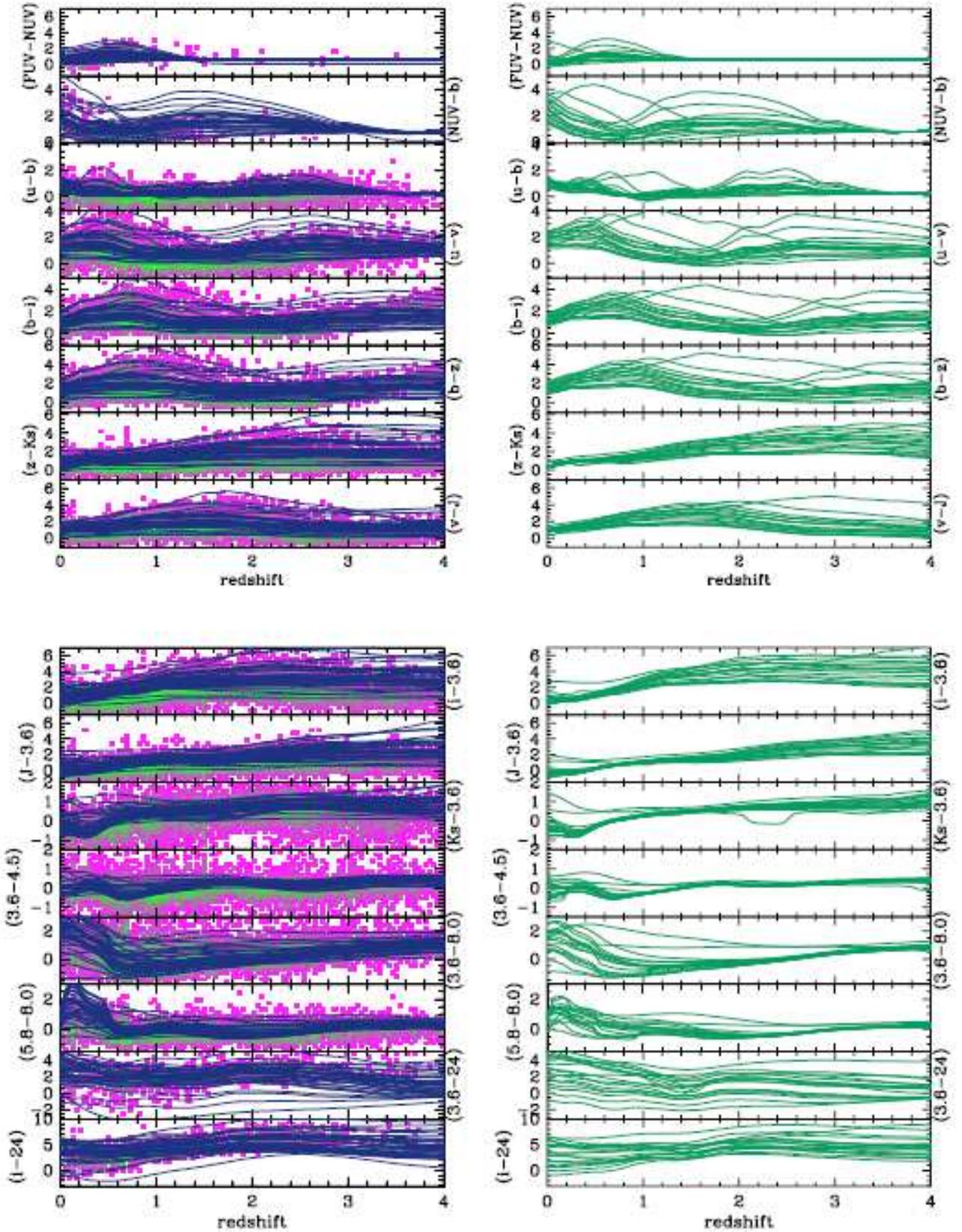}
\caption{Trends of colors as a function of redshift for the new {\sc FastEM} templates
({\em left}) and \citet{polletta2007} templates ({\em right}), compared to GOODS-S data. 
In each panel, only sources detected in the two bands needed to build the given color 
are taken into account. We plot the density of sources in each color-redshift bin, using a logarithmic
color scale ranging from magenta (lowest density) to light green (highest density).}
\label{fig:k_corr}
\end{figure*}

\addtocounter{figure}{-1}

\begin{figure*}[!ht]
\centering
\includegraphics[height=0.45\textheight]{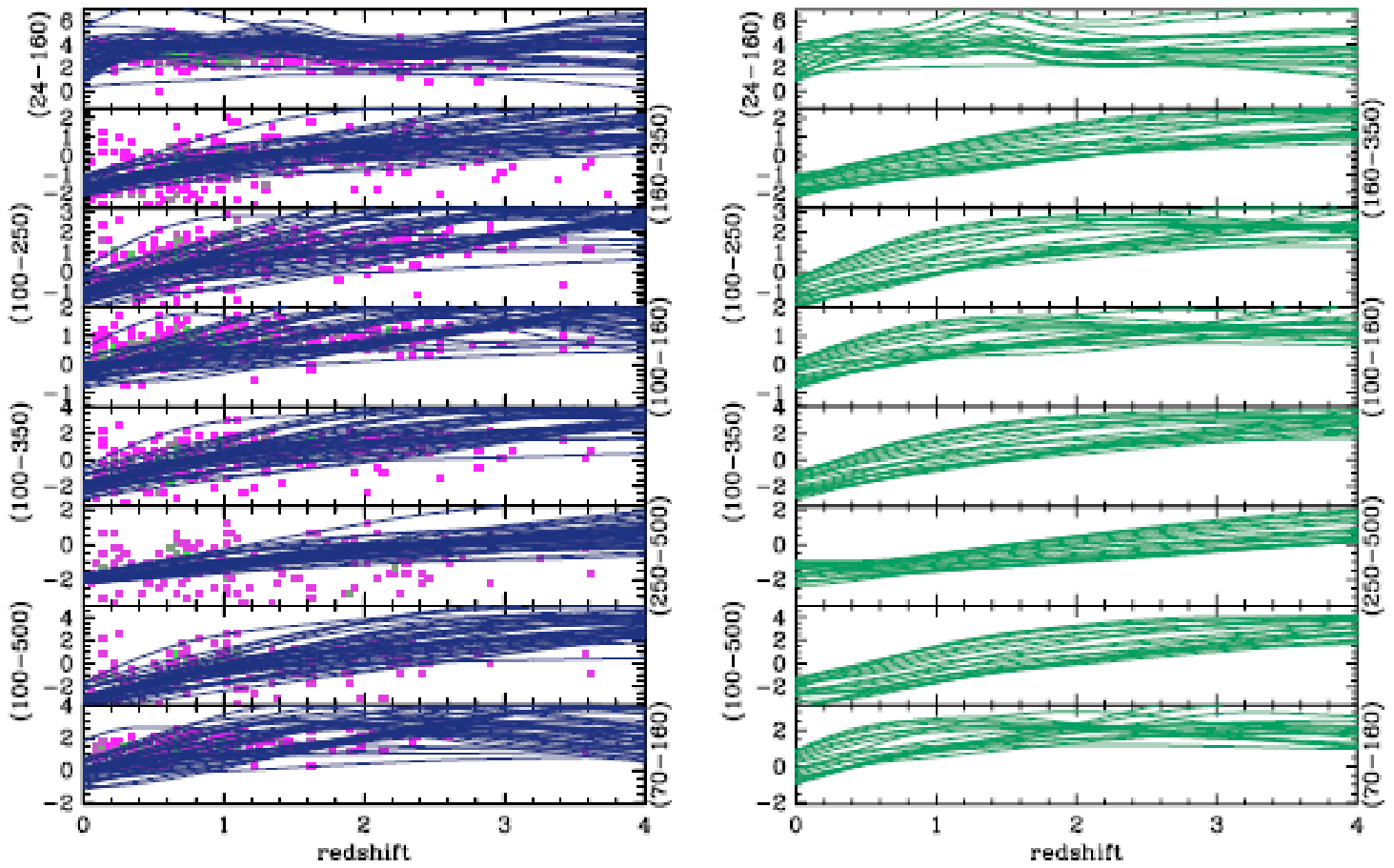}
\caption{continued.}
\end{figure*}

The newly generated SED templates were obtained 
on a limited selection of sources drawn from GOODS-N/S and COSMOS 
multi-wavelength catalogs (see Sect. \ref{sect:data}). 
Thus, it is worth comparing them against the photometry of all sources detected 
in any band.
In the left panels of Fig. \ref{fig:k_corr}, we show the color tracks 
as a function of redshift, computed for the new set of templates (dark blue, solid lines).
These are plotted onto the position of all real sources in the GOODS-S field
(roughly 18\,000 sources in total), 
with the sole requirement that a detection in the two bands giving each color 
is guaranteed. To avoid confusion due to the plethora of points, in this Figure we 
actually draw the density of objects in small color-redshift bins, logarithmically color-coded 
from magenta (lowest density) to light green.

The right-hand panels of Fig. \ref{fig:k_corr} show the same color-redshift tracks for 
the \citet{polletta2007} template set. 

The main advantage of our new library, based on actual \textit{Herschel} data, is that 
it spans a wider color range, 
with respect to the \citet{polletta2007} one, both on the blue and red
MIR-FIR color sides. The same applies also to the optical and NIR 
domains, with slight differences varying from case to case.
When comparing to observed data, the new library based on \textit{Herschel} data
covers most of the observed colors, with the exception of the bluest side of the 
NIR color space. This can be due to a combination of the large scatter in 
the IRAC observed photometry and a possible missing set 
of passive templates (remember that the library was defined on \textit{Herschel}-detected objects). 
However note that the Polletta library spans a similar NIR color range, although 
it contains a number of elliptical galaxy templates of different ages, obtained with 
GraSil \citep{silva1998}.

The SEDs of all 24 $\mu$m objects in the GOODS-N/S catalogs are then 
fit with the same B13 setup as in Sect. \ref{sect:testing}, with the purpose of classifying 
the whole population of infrared galaxies with the information provided by the 
new library. The list of sources detected by MIPS in the MIR (see Table \ref{tab:depths}) 
comprises 2575 objects in GOODS-N and 1712 in GOODS-S.

Three main parameters are examined: the fraction of infrared luminosity 
$L(8$--$1000\,\mu\textnormal{m})$ contributed by the possible AGN component, 
the $L(60)/L(100)$ restframe FIR color, and the ratio $L(160)/L(1.6)$ between 
FIR and NIR restframe luminosities.
For ease of interpretation, templates are grouped in three classes for each 
of the parameters considered: 

\begin{description}
\item{(a)} $f(\textnormal{AGN})=0.0$,\\ \ $0.0<f(\textnormal{AGN})\le0.3$,\\ \ $f(\textnormal{AGN})>0.3$; 
\item{(b)} $L(60)/L(100)\le0.5$,\\ \ $0.5<L(60)/L(100)\le1.0$,\\ \ $L(60)/L(100)>1.0$; 
\item{(c)} $L(160)/L(1.6)\le10$,\\ \ $10<L(160)/L(1.6)\le100$,\\ \ $L(160)/L(1.6)>100$.
\end{description}

\begin{figure}[!ht]
\centering
\includegraphics[width=0.45\textwidth]{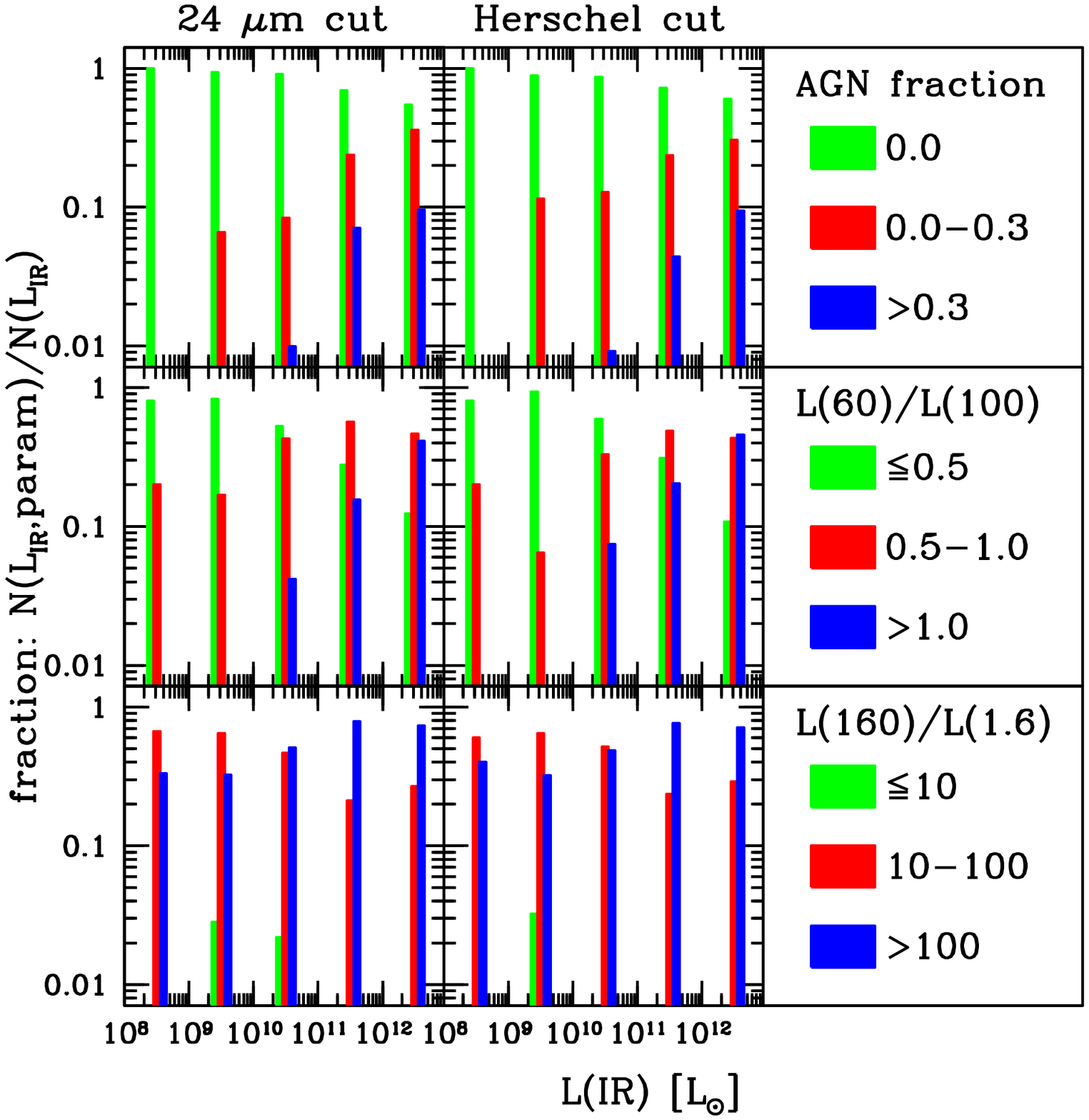}
\caption{Relative distribution of GOODS-S sources as a function of 
AGN $L(\textnormal{IR})$ fraction, $L(60)/L(100)$ and $L(160)/L(1.6)$, as indicated by the labeling in 
the right-hand panels, in different IR luminosity bins. Histograms are normalized to the number 
of sources in each $L(\textnormal{IR})$ bin. 
Left/right panels refer to a 24 $\mu$m selection and sources detected by \textit{Herschel}, respectively. Infrared luminosity bins are 1 dex wide, i.e.
they cover ranges $10^8-10^9$, $10^9-10^{10}$, etc.}
\label{fig:classification1}
\end{figure}

In what follows we analyze GOODS-S data; similar results are obtained in GOODS-N.
Figure \ref{fig:classification1} reports on the relative distribution of 
sources at all redshifts in each bin of $f(\textnormal{AGN})$, $L(60)/L(100)$ and $L(160)/L(1.6)$,
grouped in 1 dex wide bins of IR luminosity. 
When analyzing classes {\em (b)} and {\em (c)} objects hosting an AGN (on the basis of our SED fitting) 
have been excluded, because of the contamination of MIR and NIR 
fluxes by the torus component.

As infrared luminosity increases, 
the fraction of detected objects hosting an AGN with respect to the total 
number in the given $L(\textnormal{IR})$ bin tends to increase. This is particularly true 
for extreme sources, with $f(\textnormal{AGN})>0.30$. Nevertheless, their incidence over the 
whole detected sample is well below 10\%.

As far as the $L(60)/L(100)$ color is concerned, the classification based 
on SED fitting shows how the relative number of bluer objects increases
as a function of luminosity, at the expense of $L(60)/L(100)\le0.5$ sources. 
At the same time, FIR emission becomes more prominent, with 
respect to NIR, and $L(160)/L(1.6)>100$ galaxies dominate the 
bright end. Results obtained including (right hand panels) and excluding (left panels) 
\textit{Herschel} bands in the fit are remarkably similar, but one should keep in mind that 
roughly 75\% of the sources in the analysis benefit from 
a \textit{Herschel} detection. The reader is thus 
warned that results based on $L(60)/L(100)$ and 24 $\mu$m detections only (left panels) should 
be taken with care, if a higher fraction of objects is missing a FIR detection.

\begin{figure*}[!ht]
\centering
\rotatebox{-90}{\includegraphics[height=0.70\textwidth]{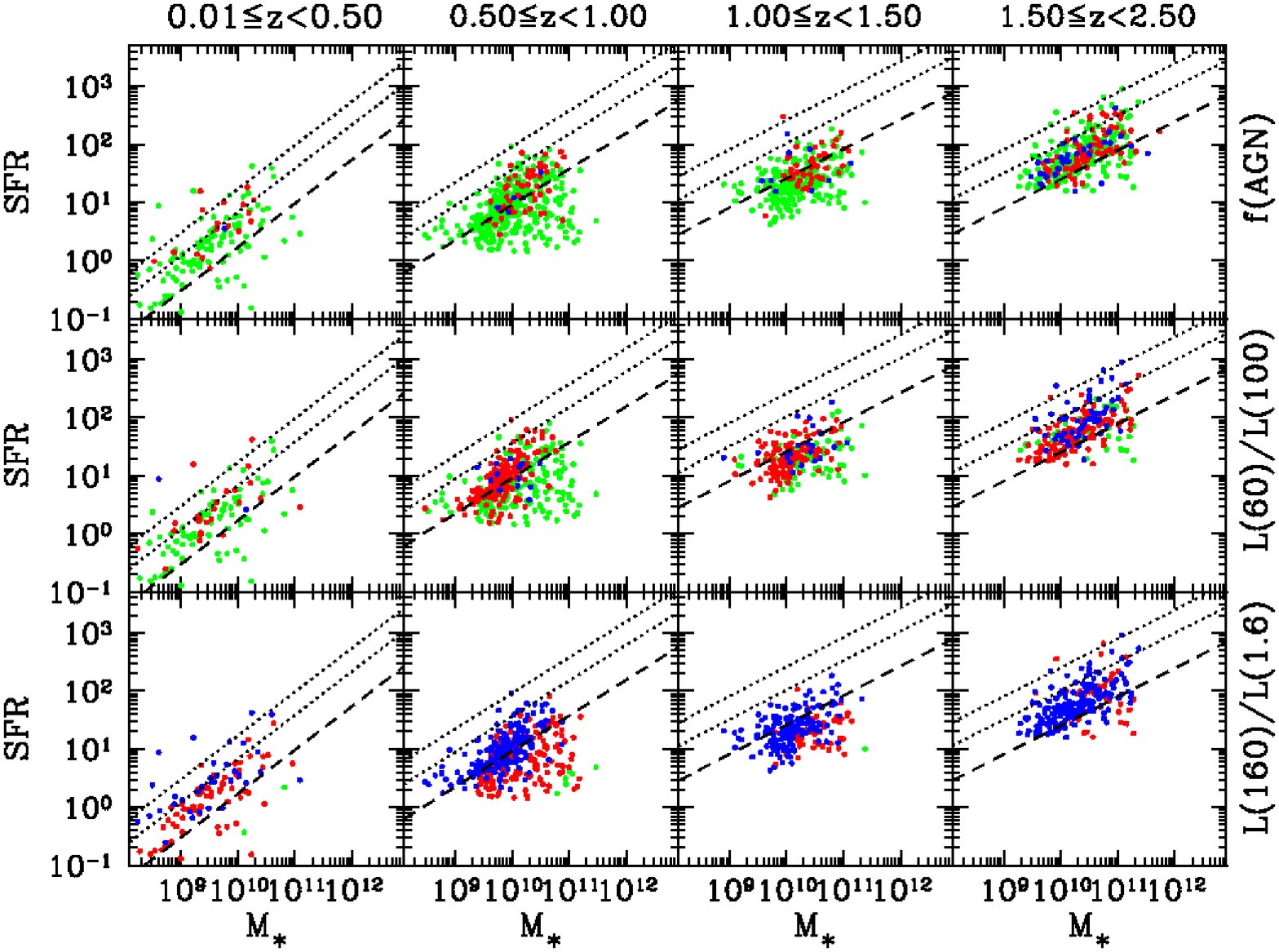}}
\caption{Position of GOODS-S 24 $\mu$m sources in the 
$M_\ast$--$\textnormal{SFR}$--$z$ space. Top, middle, and bottom panels depict objects color coded 
on the basis of $L(\textnormal{IR})$ AGN fraction, $L(60)/L(100)$ luminosity ratio and 
$L(160)/L(1.6)$ color, respectively. Color coding is the same as in 
Fig. \ref{fig:classification1}. Dashed lines represent the position of the 
``main sequence'' of star formation, as derived by \citet{rodighiero2010}
by stacking 4.5\,$\mu$m-selected sources. Dotted lines trace the MS +0.6 and +1.0 dex, respectively.}
\label{fig:m_sfr}
\end{figure*}

The $L(160)/L(1.6)$ ratio is a proxy for the specific star formation rate (sSFR), 
while $L(60)/L(100)$ has often been adopted as a tracer of dust temperature \citep[e.g.][]{dale2002}.
Figure \ref{fig:m_sfr} shows the $M_\ast$--$\textnormal{SFR}$--$z$ space 
for GOODS-S 24 $\mu$m detected sources. Star formation rates are based on 
our estimate of $L(\textnormal{IR})$ and the \citet{kennicutt1998} conversion 
to SFR. An independent estimate, obtained through a ladder of SFR tracers, and 
calibrated on SFR$_{{\rm UV+IR}}$ \citep{wuyts2011a,wuyts2011b} produces similar results. 
Stellar masses $M_\ast$ 
are based on \citet{bruzual2003} models, as applied to our data by \citet{wuyts2011b}.
Note that, if the \citet{maraston2005} models with enhanced emission by TP-AGB 
(thermally pulsating asymptotic giant branch) stars 
would have been adopted, best fits would correspond to younger, less massive (and more actively starforming)
systems than inferred with BC03 \citep{wuyts2011a,santini2009,salimbeni2009}.
When the two color classification schemes are adopted (middle and bottom panels), we exclude 
sources with an AGN component. The incidence of warmer sources 
(larger 60 to 100 $\mu$m luminosity ratio) grows for objects with larger 
FIR ``excess''. 
This is in line with what is found by Magnelli et al. (in prep.)
studying the dependence of dust temperature $T_{\rm dust}$ as a function of 
position in the $M_\ast$--$\textnormal{SFR}$--$z$ space and distance from the so-called ``main sequence'' 
of star formation, and by \citet{symeonidis2013} studying the $L$--$T$ relation 
of \textit{Herschel}-selected IR-luminous galaxies. For reference, 
the MS locus defined by stacking 4.5 $\mu$m selected galaxies \citep{rodighiero2010}
is shown by dashed lines (obtained by interpolating between Rodighiero's redshift bins),
and the levels of MS +0.6 and +1.0 dex are marked with dotted lines. 
\citet{rodighiero2011} used a +0.6 dex threshold to isolate off-sequence galaxies at $1.5<z<2.5$. 
Note that, when limiting the analysis to \textit{Herschel}-detected sources only, objects lying below the main sequence
at $z\ge0.5$ are excluded.

Segregation on the basis of $L(60)/L(100)$
or $L(160)/L(1.6)$ is clearly seen, although it becomes  
gradually diluted at larger cosmic distances, possibly because an increasing incompleteness 
affects the locus of MS galaxies (see e.g. Magnelli et al., in prep.). 
The $L(160)/L(1.6)<10$ bin is hardly sampled by MIR-FIR detections:
the few sources belonging to this class indeed turn out to lie off-sequence, in the 
locus occupied by passive galaxies in the $M_\ast-SFR$ plane, and are characterized by our 
{\em outlier} quiescent templates.
Finally, in the $M_\ast$--$\textnormal{SFR}$ diagram, above $M_\ast\sim 10^{10}$~M$_\odot$, AGN appear to be distributed along with all 
other galaxies  \citep[see also][]{santini2012}.
They have a tendency to lie on the high SFR side of the source distribution across the 
``main sequence'', but with no evident dependence on the fraction 
of infrared luminosity -- $f(AGN)$ -- they are powering.


\section{Summary}

Starting from FIR detected sources and including multi-wavelength 
information from the ultraviolet to optical, NIR and MIR, 
we have grouped a subsample of sources in the deepest and in the widest 
fields observed by \textit{Herschel} within the overlapping PEP plus HerMES dataset: GOODS-S, GOODS-N and COSMOS.
Following the work by \citet{davoodi2006}, the color distribution of 
galaxies has been modeled as the superposition of multi-variate Gaussian 
modes. The chosen parameter space, consisting of 10 restframe colors in the wavelength 
range between 1400 \AA\ and 100 $\mu$m, covers 
galaxy properties such as the UV slope, D4000 break, 1.6 $\mu$m stellar peak, 
PAH and warm dust emission, possible AGN contamination and cold dust contribution.

A modified version of the {\sc magphys} \citep{dacunha2008} code has been developed, 
aimed at reproducing galaxy SEDs with the combination 
of three emitting components: stellar emission, thermal emission from dust heated by stellar 
UV-optical radiation, and a possible AGN torus. The median SEDs of each Gaussian mode 
have been fit with this code and with the original {\sc magphys} code, thus providing 
a reliable interpolation over a continuous wavelength range. A new library of SED 
templates has thus been defined. 

The main results of this grouping and SED fitting classification are:

\begin{itemize}
\item the distribution of \textit{Herschel}-detected galaxies in the 10 restframe color space 
can be reproduced by six to nine multivariate Gaussian modes in the three fields considered, 
with some differences due to depth of data, wavelength coverage, and how fine the grid 
of observed bands is. 
\item the classification thus obtained has been used to identify rare objects, 
so-called ``outliers'' in the grouping scheme. The main classes of outliers 
identified are torus-dominated objects, type-1 AGN with featureless SEDs, and very bright 
FIR emitters. Two low redshift elliptical and one passively evolving spiral, 
well detected by \textit{Herschel}, provide the completion of the classification picture.
\item fitting the median SED of each group and outlier with multi-component SED synthesis, 
a new set of templates has been defined\footnote{The new library can be retrieved at 
http://www.mpe.mpg.de/ir/Research/PEP/uvfir\_templ, 
normalized by $L(\textnormal{IR})$, $L(1.6\ \mu{\rm m})$, and $M_\ast$ (based on the results of SED fitting).}. 
Among others, it includes five type-1 AGN models, 
and five type-2, characterized by different AGN torus contributions to total 
infrared (8-1000 $\mu$m) luminosity.
\item the new set of templates has been compared to other libraries and methods 
often adopted in the literature to derive infrared luminosities of galaxies (see Sect. \ref{sect:testing} for a list). 
When including FIR data, the $L(\textnormal{IR})$ estimates obtained 
with the different methods are consistent with each other to within 10--20\%. 
Scatter in comparing different derivations of $L(\textnormal{IR})$ strongly depends 
on the choice of codes and their fine tuning.
\item comparing $L(\textnormal{IR})$ derived with or without the inclusion of \textit{Herschel} data, all 
methods are affected by a significant scatter.
Luminosity-dependent approaches, calibrated on local galaxies, turn out to be affected by more critical problems 
\citep[e.g.][]{elbaz2010,nordon2010} than 
simple SED fitting allowing for free rescaling, but appropriate corrections account for 
most systematics \citep[e.g.][]{elbaz2011,nordon2012}.
\item when compared to full catalogs, independently from any selection cut, 
the new templates cover most of the color range occupied by real sources at all wavelengths.
\item 24 $\mu$m sources in the two GOODS fields have been classified on the basis 
of their $L(\textnormal{IR})$ AGN fraction, $L(60)/L(100)$ color and $L(160)/L(1.6)$ flux 
ratio, as derived by SED fitting with our new library.
The incidence of warmer sources grows for objects with higher 
sSFR, in line with what is found by Magnelli et al. 
(in prep.) through a detailed study of $T_{\rm dust}$ 
in the $M_\ast$--$\textnormal{SFR}$ plane, and with results on the $L-T$ relation 
of IR SEDs by \citet{symeonidis2013}. 
AGN appear to be distributed along with all 
other galaxies, above $M_\ast\sim 10^{10}$ M$_\odot$, 
with the tendency to lie on the high SFR side of the 
``main sequence'', but with no evident dependence on the fraction 
of infrared luminosity they are powering. 
\end{itemize}

The new library of semi-empirical SEDs describes 
the actual scatter of FIR detected sources in color space, from the UV to 
the FIR itself. Direct applications can span from deriving the 
contribution of AGN and various populations to mid and FIR luminosity functions 
and $L(\textnormal{IR})$ or SFR density as a function of look-back time \citep[see also][]{gruppioni2013}, 
to describing the SEDs of individual sources.


\begin{acknowledgements}  
The authors wish to thank Dr. Jacopo Fritz for providing his
library of AGN torus emission models, and the anonymous referee for 
her/his useful comments.
This work made use of the code {\sc FastEM}, distributed by the Auton Lab, and developed by 
Andrew Moore, Paul Hsiung, Peter Sand, point of contact Saswati Ray.
PACS has been developed by a consortium of institutes led by MPE (Germany) and 
including UVIE (Austria); KU Leuven, CSL, IMEC (Belgium); CEA, LAM (France); 
MPIA (Germany); INAF-IFSI/OAA/OAP/OAT, LENS, SISSA (Italy); IAC (Spain). 
This development has been supported by the funding agencies BMVIT (Austria), 
ESA-PRODEX (Belgium), CEA/CNES (France), DLR (Germany), ASI/INAF (Italy), 
and CICYT/MCYT (Spain). SPIRE has been developed by a consortium of institutes
led by Cardiff University (UK) and including University of Lethbridge (Canada),
NAOC (China), CEA, LAM (France), IFSI, University of Padua (Italy), IAC
(Spain), Stockholm Observatory (Sweden), Imperial College London, RAL,
UCL-MSSL, UKATC, University of Sussex (UK), Caltech, JPL, NHSC,
University of Colorado (USA). This development has been supported by national
funding agencies: CSA (Canada); NAOC (China); CEA, CNES, CNRS (France);
ASI (Italy); MCINN (Spain); SNSB (Sweden); STFC, UKSA (UK) and NASA
(USA).
\end{acknowledgements}




\bibliographystyle{aa}
\bibliography{b2013}



\begin{appendix} 

\section{Results of SED fitting}\label{sect:sed_fit_imgs}

In this Appendix, the results of SED fitting are presented. 
For each median SED or outlier, only the best fit solution 
is shown, either with or without the AGN component added. 

We recall (see Sect. \ref{sect:sed_fit}) that the {\sc magphys} 
code \citep{dacunha2008} was adopted to reproduce SEDs. 
It combines BC03 optical/NIR stellar models, including 
the effects of dust attenuation as prescribed by \citet{charlot2000}, to MIR/FIR dust emission
computed as in \citet{dacunha2008}, linking the two components 
through energy balance: the total energy absorbed by dust in stellar birth clouds 
and in the ambient interstellar medium is re-distributed at infrared wavelengths. 

The code has then been modified to include the possible 
contribution of an AGN torus component, using the \citet{fritz2006}
library, in order to overcome one of {\sc magphys} main assumptions, 
implying that the only source of dust heating is starlight.

Figure \ref{fig:sed_fit1}
shows our best fit solutions (i.e. those with minimum $\chi^2$). Table \ref{tab:sed_fits} summarizes results, 
including a brief description of each template, restframe colors spanning from the 
$u$ band to 160 $\mu$m, and AGN contribution to the infrared (8--1000 $\mu$m) 
luminosity. The latter is reported only for objects that require an AGN component, 
while sources best reproduced by the original {\sc magphys} code have a $L(\textrm{IR})$ 
AGN fraction $<$1\%.

Restframe colors have been obtained by convolving best fit models with 
filter transmission curves.
Descriptions provided in Table \ref{tab:sed_fits} highlight the main features 
of templates, e.g. optical colors (red/blue), position of the FIR peak (warm/cold),
intensity of PAH emission, optical extinction, AGN contribution, as well as 
additional known properties of individual outliers. Type-1 and Type-2 AGN labels
refer simply to the best fit solution and are defined such that in ``type-2'' models 
the line of sight intersects the torus dust distribution 
($\Phi>90^\circ-\Theta/2$, with $\Phi$ defined starting from the polar axis, 
see Fritz et al. \citeyear{fritz2006}), and vice versa in ``type-1''
cases.

\begin{landscape}
\begin{table}
\tiny
\centering
\begin{tabular}{l c l | c c c c c c c c | c c c}
\hline
\hline
\multicolumn{3}{c|}{Template}                       & \multicolumn{8}{c|}{Colors} & \multicolumn{3}{c}{$L(\textnormal{IR})$ AGN fraction}\\
Name &	ID  		     & Description & $(u-b)$ & $(b-r)$ & $(r-K_{\rm s})$ & $(K_{\rm s}-4.5)$ & $(r-24)$ & $(24-100)$ & $(H-160)$ & $(60-100)$ & best & median  & range$^\dagger$     \\
\hline
Red-SF-glx-1     & 01\_01  & Red star-forming galaxy       & 0.80 & 0.77 &  1.32 & -0.51 &  5.84 & 2.08 & 5.95 & -0.05 &  --  &  --  & --	    \\   
Secular-glx      & 01\_02  & Secularly evolving galaxy     & 1.04 & 0.89 &  1.05 & -1.11 &  1.38 & 3.72 & 3.44 &  0.84 &  --  &  --  & --	    \\   
Mod-SF-glx       & 01\_03  & Moderate star-forming galaxy  & 0.93 & 0.51 &  0.97 & -0.92 &  1.87 & 4.39 & 4.85 &  0.75 &  --  &  --  & --	    \\   
Young-SF-glx     & 01\_04  & Young star-forming galaxy     & 0.21 & 0.16 &  0.52 & -0.09 &  3.65 & 4.84 & 7.57 &  1.11 &  --  &  --  & --	    \\   
WeakPAH-SF-glx-1 & 01\_05  & Weak-PAH star-forming galaxy  & 0.88 & 0.43 &  1.02 & -0.83 &  3.28 & 3.31 & 5.18 &  0.75 &  --  &  --  & --	    \\   
Type2-AGN-1      & 01\_06  & Type-2 AGN		           & 1.15 & 0.92 &  1.92 &  1.40 &  6.48 & 2.64 & 7.17 &  0.62 & 0.25 & 0.30 & 0.17-0.53       \\   
\hline
Warm-SF-glx	& 02\_01  & Warm star-forming galaxy	 & 0.67 & 0.92 &  1.71 & -0.32 &  6.75 & 1.79 & 6.32 & -0.16 &  --  &  --  & -- 	    \\
Type1-AGN-1	& 02\_02  & Type-1 AGN		  	 & 0.61 & 0.32 &  1.47 &  0.54 &  3.12 & 0.28 & 2.23 & -0.11 & 0.72 & 0.61 & 0.53-0.79    \\
SF-glx-1	& 02\_03  & Star-forming galaxy	  	 & 0.95 & 0.50 &  0.91 & -0.96 &  1.88 & 4.06 & 4.84 &  1.30 &  --  &  --  & -- 	    \\
Type2-AGN-2	& 02\_04  & Type-2 AGN		  	 & 1.10 & 1.04 &  1.41 &  0.40 &  5.92 & 2.37 & 6.34 &  0.71 & 0.33 & 0.20 & 0.12-0.56    \\
SF-Type2-AGN-1	& 02\_05  & Warm, star-forming, host-dom, type-2 AGN  & 0.91 & 0.54 &  1.14 &  0.07 &  3.99 & 2.27 & 5.06 &  0.27 & 0.16 & 0.11 & $\le$0.20    \\
Obs-SF-glx	& 02\_06  & Obscured star-forming galaxy  	   & 1.10 & 1.04 &  1.42 & -0.69 &  5.49 & 1.61 & 4.75 & -0.39 &  --  &  --  & --	    \\
MIRex-SF-glx	& 02\_07  & MIR-excess star-forming galaxy	   & 0.80 & 0.31 &  0.82 & -0.80 &  4.12 & 2.50 & 5.40 &  0.41 &  --  &  --  & --	    \\
Cold-glx	& 02\_08  & Cold, secularly evolving galaxy	   & 1.09 & 0.78 &  1.03 & -1.09 & -0.01 & 5.15 & 4.27 &  1.07 &  --  &  --  & --	    \\
Red-SF-glx-2	& 02\_09  & Red star-forming galaxy		   & 1.39 & 1.53 &  1.85 & -0.77 &  4.43 & 3.62 & 5.36 &  0.08 &  --  &  --  & --	    \\
\hline
SF-Type1-AGN-1    & 03\_01  & Star-forming type-1 AGN		           & 0.32 & 0.45 &  1.26 &  0.35 &  3.60 & 1.96 & 4.66 &  0.72 & 0.33 & 0.30 & 0.24-0.47	\\
SF-Type2-AGN-2    & 03\_02  & Red, host-dom. star-forming type-2 AGN      & 1.11 & 0.85 &  1.52 & -0.06 &  3.79 & 3.66 & 5.83 &  0.48 & 0.04 & 0.04 & $\le$0.05	\\
SF-Type2-AGN-3    & 03\_03  & Blue, host-dom. star-forming type-2 AGN     & 0.84 & 0.77 &  1.20 &  0.21 &  4.29 & 3.34 & 6.15 &  0.45 & 0.02 & 0.03 & $\le$0.08	\\
PAH-SF-glx        & 03\_04  & Enhanced-PAH star-forming galaxy            & 0.68 & 0.36 &  1.04 &  0.13 &  5.56 & 2.25 & 6.65 &  0.19 &  --  &  --  & --	    \\
WeakPAH-SF-glx-2  & 03\_05  & Weak-PAH star-forming galaxy 	           & 0.99 & 0.81 &  0.68 & -0.92 &  2.33 & 3.92 & 5.20 &  0.90 &  --  &  --  & --	    \\
BroadFIR-SF-glx   & 03\_06  & Broad-FIR peak star-forming galaxy           & 0.76 & 0.61 &  0.73 & -0.63 &  3.56 & 3.33 & 5.89 &  0.38 &  --  &  --  & --	    \\
MIR-powlaw-SF-glx & 03\_07  & 3 $\mu$m power-law, star-forming type-2 AGN  & 0.80 & 0.88 &  1.60 &  1.44 &  5.81 & 2.26 & 5.99 &  0.03 & 0.12 & 0.28 & 0.10-0.61	\\
SF-glx-2          & 03\_08  & Star-forming galaxy                          & 0.85 & 0.46 &  1.07 & -0.69 &  3.28 & 4.04 & 6.12 &  0.49 &  --  &  --  & --	    \\
Blue-SF-glx       & 03\_09  & Blue, MIR-excess star-forming galaxy         & 0.81 & 0.35 &  0.92 & -0.56 &  4.11 & 2.60 & 5.52 &  0.50 &  --  &  --  & --	    \\
\hline
Torus		& 04\_01  & Torus-dominated galaxy		& 1.40 & 1.56 &  3.04 &  1.07 &  5.76 & 1.61 & 4.70 &  0.56 & 0.54 & 0.42 & 0.35-0.62 \\      
SF-Type1-AGN-2 	& 04\_02  & Blue, star-forming, type-1 AGN	& 0.31 & 0.16 &  1.77 &  0.24 &  4.54 & 1.96 & 4.67 &  0.17 & 0.05 & 0.14 & 0.04-0.20 \\      
SF-Type1-AGN-3	& 04\_03  & Red, star-forming, type-1 AGN	& 0.54 & 0.47 &  2.03 &  0.29 &  3.90 & 3.36 & 5.05 &  0.52 & 0.11 & 0.09 & 0.05-0.25 \\      
SF-Type1-AGN-4  & 04\_04  & Host-dom., star-forming type-1 AGN 	& 0.59 & 0.85 &  1.94 &  0.86 &  6.26 & 3.33 & 7.72 &  0.16 & 0.01 & 0.06 & 0.01-0.25 \\      
Si-break   	& 04\_05  & Silicate break galaxy		& 1.19 & 1.05 &  2.14 & -0.47 &  6.28 & 2.56 & 6.39 &  0.33 & --   &  --  & --        \\      
Ly-break   	& 04\_06  & Ly-break galaxy			& 0.36 & 0.08 & -0.05 & -0.16 &  3.45 & 3.98 & 6.85 &  0.66 & --   &  --  & --        \\      
Spiral     	& 04\_07  & Secularly evolving spiral	  	& 1.07 & 1.04 &  0.54 & -1.29 & -0.38 & 4.93 & 4.32 &  1.95 & --   &  --  & --        \\      
Elliptical 	& 04\_08  & Elliptical Galaxy			& 1.46 & 1.38 &  1.13 & -1.21 & -0.72 & 4.07 & 1.80 &  1.22 & --   &  --  & --        \\      
\hline
\multicolumn{14}{l}{$^\dagger$ The range of AGN fraction is computed between the 2.5$^{\rm th}$ and 97.5$^{\rm th}$ percentiles of the probability distribution function.}\\
\end{tabular}
\caption{Description of new templates, including colors $[$mag$]$ in the optical to FIR wavelength range and 
the contribution of a possible AGN to the 8-1000 $\mu$m luminosity.}
\label{tab:sed_fits}
\end{table}
\end{landscape}

\begin{figure*}[!ht]
\centering
\includegraphics[width=0.96\textwidth]{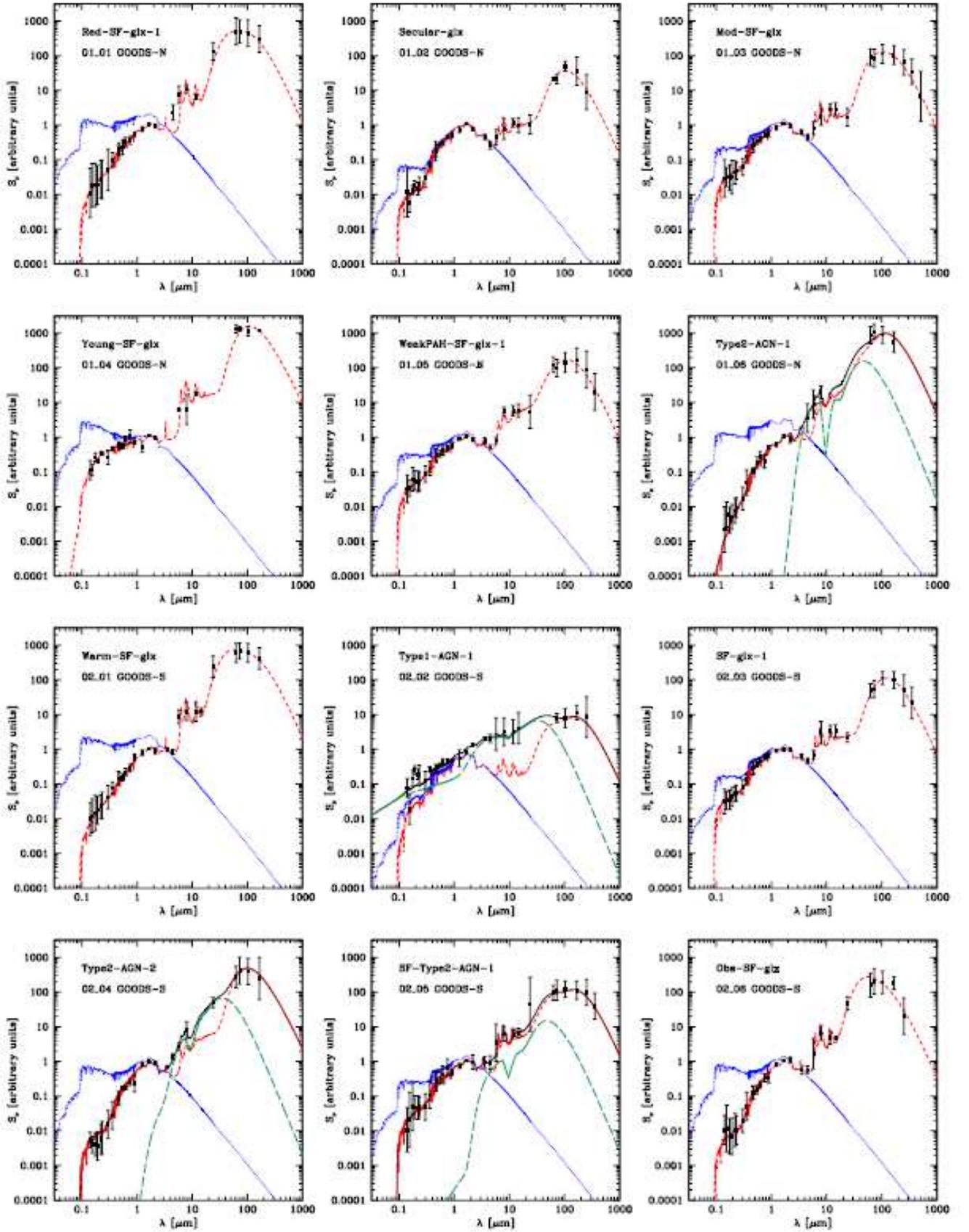}
\caption{Best fit to median SEDs, as obtained either with the {\sc magphys} code \citep{dacunha2008}
or with the modified version including a torus component \citep[using the][library]{fritz2006}.
Blue dotted lines represent the un-absorbed stellar component, while red dashed lines 
trace the combination of extinguished stars and dust infrared emission. The torus 
component is depicted with long-dashed green lines, and black solid lines are
the total emission, in those cases for which an AGN is needed.}
\label{fig:sed_fit1}
\end{figure*}

\addtocounter{figure}{-1}

\begin{figure*}[!ht]
\centering
\includegraphics[width=0.96\textwidth]{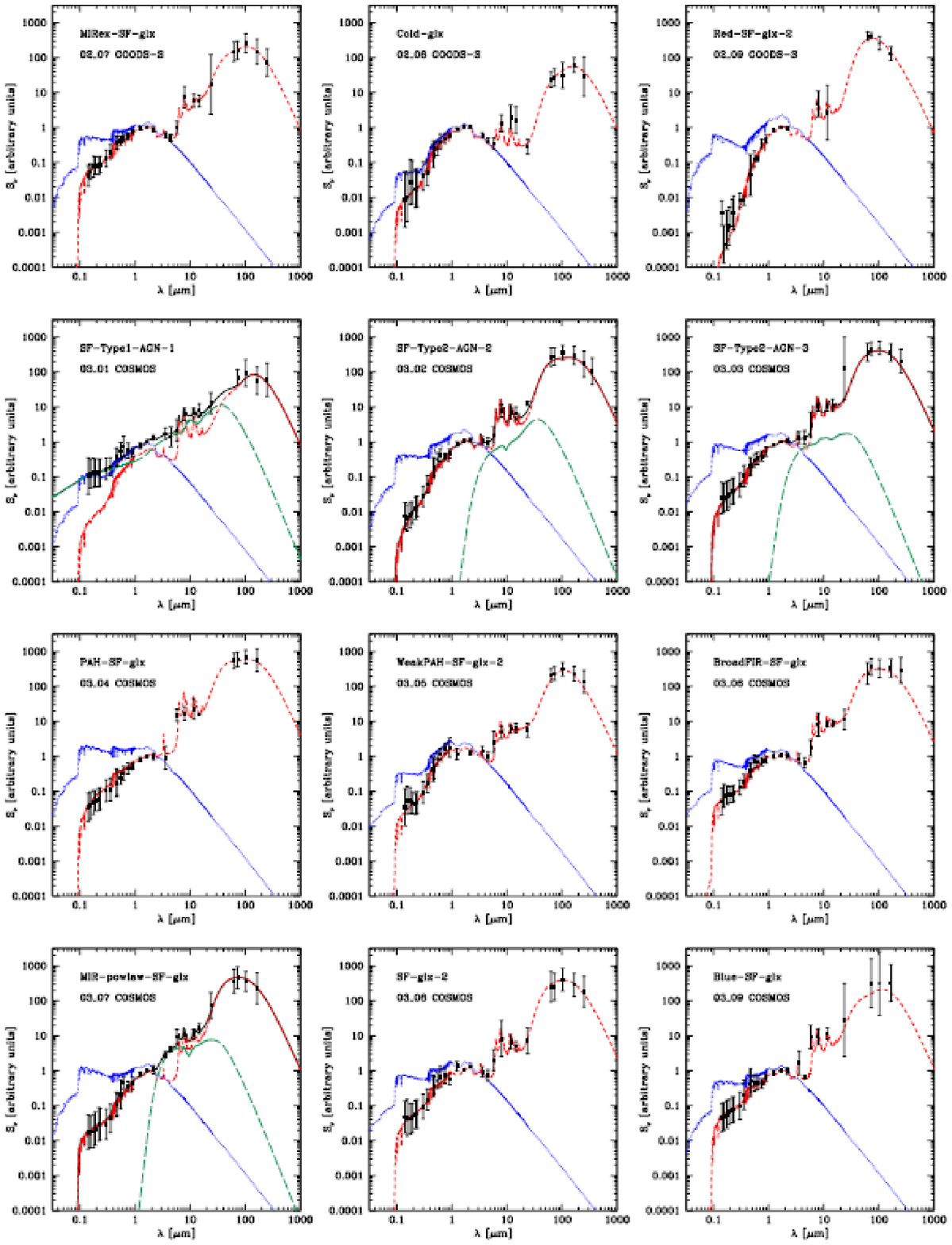}
\caption{continued.}
\end{figure*}

\addtocounter{figure}{-1}

\begin{figure*}[!ht]
\centering
\includegraphics[width=0.96\textwidth]{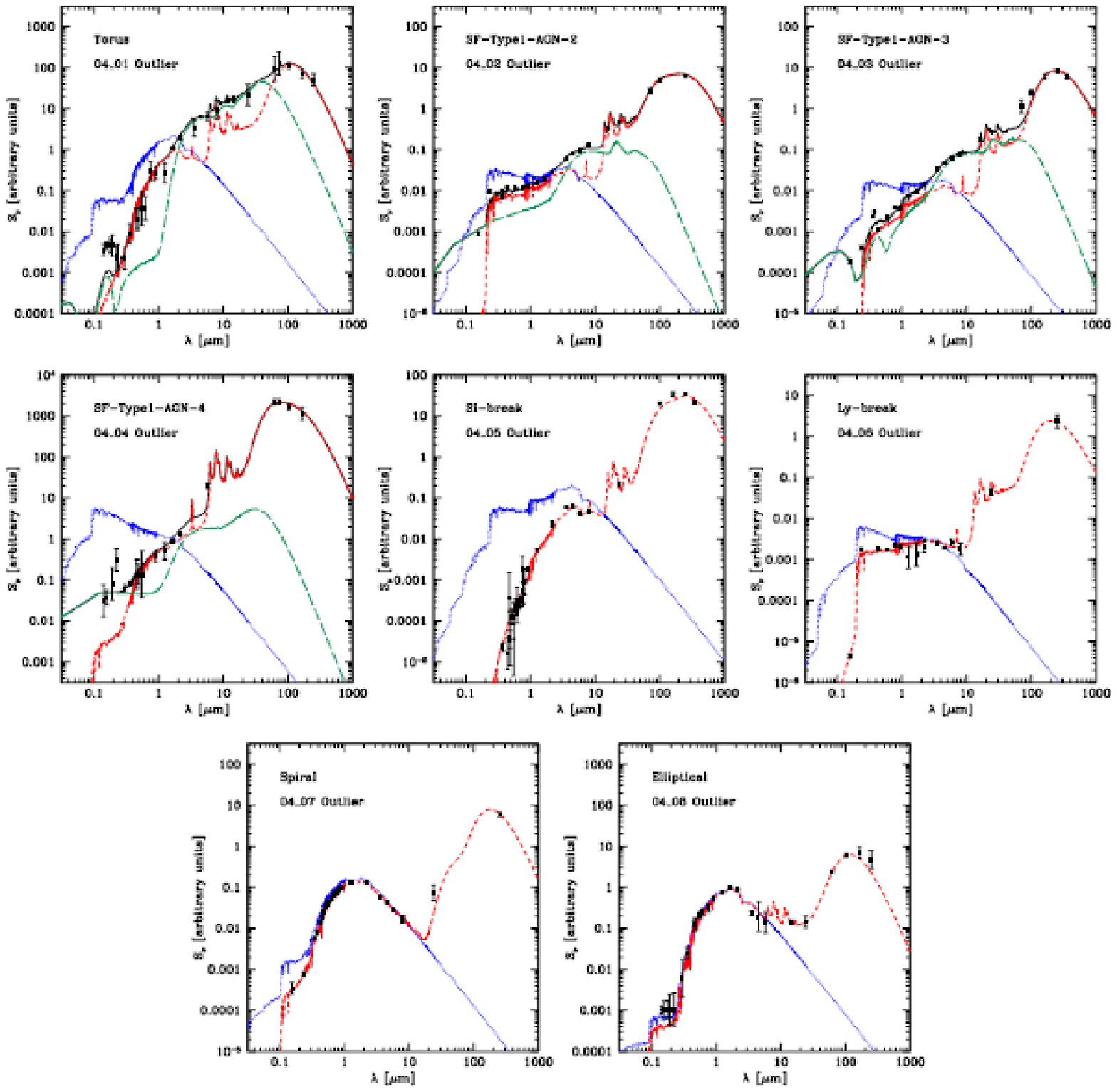}
\caption{continued.}
\end{figure*}

\end{appendix}


\end{document}